\documentclass[prd,showkeys,amsmath,amssymb,twocolumn,superscriptaddress,nofootinbib,showpacs]{revtex4}
\usepackage{mathrsfs}
\usepackage{graphicx,subfigure}
\usepackage{hyperref}
\usepackage{color}

\newcommand{\comment}[1]{}

\newcommand{\scri}{\mathscr{I}}
\newcommand{\hor}{\mathscr{H}}

\newcommand{\scrip}{{$\mathscr{I}^+$}}
\newcommand{\scribub}{{$\mathscr{H}_C^+$}}
\newcommand{\scrim}{{$\mathscr{I}^-$}}
\newcommand{\hp}{{$\mathscr{H}_1^+$}}
\newcommand{\hm}{{$\mathscr{H}_2^-$}}

\begin{document}

\title{Semiclassical instability of dynamical warp drives}
%Superluminal Warp Drives are unstable}

\date{\today}

\author{Stefano Finazzi}
\email{finazzi@sissa.it}
\author{Stefano Liberati}
\email{liberati@sissa.it}
\affiliation{SISSA, via Beirut 2-4, 34151 Trieste, Italy and INFN sezione di Trieste, via Valerio 2, 34127 Trieste, Italy.}
\author{Carlos Barcel\'o}
\email{carlos@iaa.es}
\affiliation{Instituto de Astrof\'{\i}sica de Andaluc\'{\i}a, CSIC, Camino Bajo de Hu\'etor 50, 18008 Granada, Spain}
%\thanks{finazzi@sissa.it, liberati@sissa.it}~
%International School for Advanced Studies, via Beirut 2-4, Trieste
%34014, Italy;}\\ {\small \it INFN sezione di Trieste.}\\ 
%\author{Carlos Barcel\`o}\\
%\thanks{barcelo@iass.es}~Instituto de Astrof\'{\i}sica de Andaluc\'{\i}a, CSIC, Camino Bajo de Hu\'etor 50, 18008 Granada, Spain}

%\email{finazzi@sissa.it, liberati@sissa.it, barcelo@iaa.es}

\begin{abstract}

Warp drives are very interesting configurations in general relativity: At least theoretically, they provide a way to travel at superluminal speeds,  {albeit} at the cost of requiring exotic matter to exist as solutions of Einstein's equations. {However,} even if one succeeded in providing the necessary exotic matter to build them, it would still be necessary to check whether they would survive to the switching on of quantum effects. Semiclassical corrections to warp-drive geometries have been analyzed only for eternal warp-drive bubbles traveling at fixed superluminal speeds. Here, we investigate the more  { realistic} case in which a superluminal warp drive is created out of an initially flat spacetime. First of all we analyze the causal structure of eternal and dynamical warp-drive spacetimes. Then we pass to the analysis of the renormalized stress-energy tensor (RSET) of a quantum field in these geometries. While the behavior of the RSET in these geometries has close similarities to that in the geometries associated with gravitational collapse, it shows dramatic differences too.  {On one side, an observer located at the center of a superluminal warp-drive bubble would generically experience a thermal flux of Hawking particles. On the other side, such Hawking flux will be generically extremely high if the exotic matter supporting the warp drive has its origin in a quantum field satisfying some form of quantum inequalities. Most of all, we find that the RSET will exponentially grow in time close to, and on, the front wall of the superluminal bubble. Consequently, one is led to conclude that the warp-drive geometries are unstable against semiclassical backreaction.}

%Warp drives are very interesting solutions of general relativity requiring exotic matter. So far, much attention has been devoted to the stability of stationary solutions corresponding to warp-drive bubbles traveling at superluminal speeds. However, a detailed semiclassical analysis of the onset of a superluminal warp drive starting from subluminal speeds is still missing.  Here we fill this gap by investigating the causal structure of a dynamical warp drive as well as the behavior of the renormalized stress-energy tensor of quantum fields in its geometry. While this analysis has close similarities to the case of a semiclassical gravitational collapse, it shows nonetheless dramatic behaviors peculiar to the warp-drive geometry. In particular, we find that an observer located at the center of a superluminal warp-drive bubble would generically experience a thermal bath at  {an extremely high} Hawking temperature. Furthermore, the renormalized stress-energy tensor of a scalar field will exponentially grow in time close to the front wall of the superluminal bubble, hence strongly supporting the conclusion that the warp-drive geometry is unstable against semiclassical back-reaction.
\end{abstract}


\maketitle

\section{Introduction}\label{sect:intro}

{Since Alcubierre introduced them \cite{alcubierre}, warp drives have been certainly one of the most studied spacetime geometries among those requiring exotic matter for its existence (see \cite{lobo2007} for a recent review about ``exotic spacetimes'').}
%Since they have been introduced by \cite{alcubierre}, warp-drive spacetimes have been certainly one of the most studied solutions of the Einstein equations among those requiring exotic matter (see \cite{lobo2007} for a recent review about ``exotic spacetimes''). 
{They were
%have 
immediately 
%been 
recognized as interesting configurations
%objects 
for two main reasons}. {First}, they 
%might be 
{provide}, at least theoretically, a way to travel at superluminal speeds:
{A warp drive can be described as a spheroidal bubble separating an almost flat internal region from an external
asymptotically flat spacetime with respect to which the bubble moves at an arbitrary speed. The idea is that although
on top of a spacetime nothing can move with speeds greater than that of light, spacetime itself has no \emph{a priori} restriction on the speed with which it can be stretched.}
%as they describe a bubble containing an almost flat region, moving at arbitrary speed within an asymptotically flat spacetime. 
{Second}, they 
%are 
{provide an} exciting 
%test field 
{ground for testing} our comprehension of general relativity (GR) and quantum field theory in curved spacetimes (for instance when investigating warp-drive implications for causality \cite{everett}).

In its original form \citep{alcubierre} the warp-drive geometries
%has 
{are described by} the simple expression
\begin{equation}\label{eq:3Dalcubierre}
 ds^2=-c^2 dt^2+\left[dx-v(r)dt\right]^2+dy^2+dz^2~,
\end{equation}
where $r\equiv \sqrt{(x-v_0 t)^2+y^2+z^2}$ is the distance from the center of the bubble {and} $v_0$ the warp-drive velocity. Here and thereafter $v=v_0 f(r)$ {with} $f$ a suitable smooth function satisfying $f(0)=1$ and $f(r)\to0$ for $r\to\infty$.

After the warp-drive spacetimes were proposed, the most investigated aspect of them has been the amount of exotic matter (i.e.~energy-conditions-violating matter) that would be required to support them.\footnote{While initially it was supposed that exotic matter was needed only for superluminal warp drives ($v_0>c$), it was later recognized \cite{lobovisser2004a,lobovisser2004b} that energy-conditions-violating matter is needed also for subluminal speeds. This points out that the need of exotic matter is peculiar of the warp-drive geometry itself, not appearing only in the superluminal regime.} It was soon realized that this was not only related to the size of the warp-drive bubble but also determined by the thickness of the bubble walls~\cite{pfenningford}. It was found that, if the exoticity was provided by the quantum nature of a field, satisfying therefore the so-called quantum inequalities (QI),\footnote{See~\cite{roman} for a review about quantum inequalities applied to some exotic spacetimes.} then the violations of the energy conditions would have to be confined to Planck-size regions, making the bubble-wall thickness $\Delta$ to be, accordingly, of Planck size [$\Delta \leq 10^2\,(v_0/c)\,\,L_{P}$, where $L_P$ is the Planck length]. However, it can be shown that very thin walls require very large amounts of exotic matter: e.g., in order to support a warp-drive bubble with a size of about $100$ m 
 {and}  propagating at $v_0\approx c$, one would need a total negative energy 
%$|E| \gg \,M_{\rm MilkyWay}$
$|E| \gg \,10^{11} M_\odot$.\footnote{If one could somehow avoid the QI, then it would be possible to built warp drives with much larger wall thickness. For example, for $\Delta\simeq 1$~m,  {one would only need} $|E|\gtrsim1/4\,M_\odot$.}
Perspectives for warp-drive engineering can be improved by resorting to a modified warp-drive configuration with a reduced surface area but the same bubble volume~\cite{broeck}. 
%For such solutions the 
{The} total amount of negative energy required to support 
%the 
{these} warp drives becomes quite small (for example, $|E_-|\approx 0.3 M_\odot$ for a $100$ m-radius bubble, although one has to add as well some positive energy outside the bubble $E_+\approx 2.5 M_\odot$), bringing the warp drive closer to a realistic solution albeit still far from foreseeable realizations.
%However the requirement on the wall thickness remained. Even though focusing more on other exotic spacetimes than on warp drives, \cite{krasnikov2003} explored possible ways out to QI bounds. He stressed that the result obtained using a semiclassical approach to get the amount of negative energy on very small scales may not be physically meaningful for quantum fluctuation having important effects. 

%For what regards the feasibility of warp-drive configurations, a parallel line of research has focussed on the study of their stability assuming that a warp-drive bubble can be created in some way. 
{Regarding the feasibility of warp-drive configurations, a parallel line of research has focused on the study of their robustness
against the introduction of quantum corrections to GR. In particular, in \cite{hiscock} it was studied what would be the effect of having semiclassical corrections in the case of an eternal superluminal warp drive.}
%In particular, it was studied in \cite{hiscock} the case of an eternal superluminal warp drive by discussing its stability against quantum effects. 
There, it was noticed that to an observer within the warp-drive bubble, the backward and forward walls (along the direction of motion) {look, respectively, like the horizon of a black hole and of a white hole}.  
%look respectively as the future and past event horizon of a black hole. 
By imposing over the spacetime a quantum state which is vacuum at the null infinities (i.e., what one may call the analog of the Boulware state for an eternal black hole) it was found that the renormalized stress-energy tensor (RSET) 
%had to diverge 
{diverges} at the horizons\comment {\citep[see][for a different point of view]{gonzales2000}}.\footnote{Throughout this paper we shall work in the Heisenberg representation so that only operators, not the states, evolve in time.}
{Independently of the availability of exotic matter to build the warp drive in the first place, the existence of a divergence of the RSET at the horizons would be telling us that it is not possible to create a warp-drive geometry within the context of semiclassical GR: Semiclassical effects would destroy any superluminal warp drive.}
However, in a more realistic situation, a warp drive would have to be created at a very low velocity in a given reference frame and then accelerated to superluminal speeds. One may then expect that the quantum state globally defined on such dynamical geometry would be automatically selected by the dynamics once suitable boundary conditions are provided (e.g., at early times). This is indeed the case for a gravitational collapse where it can be shown that, whenever a trapping horizon forms, the globally defined quantum state that is vacuum on \scrim~has to be thermal at \scrip~(at the Hawking temperature) and regular at the horizon. 
%{\em I.e.} one necessarily ends up with the 
%generalization 
% {analogous} for collapsing configurations of the Unruh vacuum defined on eternal black holes.
{In other words, the dynamics of the collapse avoids selecting a Boulware-like state, with its associated 
divergence at the horizon, ending up instead selecting the analogous for collapsing configurations of the Unruh vacuum state defined on eternal black holes, which leads to a perfectly regular RSET. Is the dynamics of the creation
of a warp drive, with its associated selection of the global vacuum state, able to avoid the presence of divergences
in the RSET?}
%~\footnote{
Indeed, in \cite{hiscock} it was  {already} noticed that an 
%Boulware-like state should not be expected to describe the quantum state characterizing a superluminal ward drive creation but rather an Unruh-like one.}. 
{Unruh-like state rather than a Boulware-like state should be expected to describe the quantum state characterizing a superluminal warp-drive creation}.
%}.

In this paper we want to settle this issue by explicitly considering the case of a warp drive which is created with zero velocity at early times and then accelerated up to some superluminal speed in a finite amount of time. This can be viewed as the warp-drive analog of a semiclassical black hole collapse \cite{stresstensor}. {By restricting attention to warp drives in} $1+1$ dimensions (since this is the only case for which one can carry out a complete analytic treatment), {we have} calculated the RSET in the warp-drive bubble.  {As we were expecting, we find} that in the center of the bubble there is a thermal flux  {of particles} at the Hawking temperature corresponding to the surface gravity of the {black horizon}. However, the surface gravity can be shown to be inversely proportional to the thickness of the bubble walls which, as said, has to be of  {the} order of the Planck length if  {some form of} QI holds. Hence, one has to conclude that an internal observer would soon find itself in the uncomfortable condition of being swamped by a thermal flux at the Planck temperature. Even worse, 
%while we do not find any divergence on the horizons at {any} finite time (in agreement with the Fulling-Sweeny-Wald theorem~\cite{fsw}), 
we do show that the RSET does increase exponentially with time on the white horizon {and close to it} (while it is regular {and small} on the black one). This clearly implies that warp drives {would become} rapidly unstable once superluminal speeds are reached. 

The plan of the paper is the following.  In Sec.~\ref{sect:causal} we study the causal structure of both an eternal and a dynamic warp drive. In Sec.~\ref{sect:radiation} we discuss the propagation of light rays in a dynamical warp-drive geometry in close analogy with the standard treatment for black holes.\footnote{A similar result was found by Gonz{\'a}lez-D{\'{\i}}az~\cite{gonzales2007}.} Finally, in Sec.~\ref{sect:RSET} we calculate the full RSET using the technique adopted in \cite{stresstensor} for black hole formation and look for its divergences. A summary of our results is given in Sec.~\ref{sect:conclusion}.

%-------------------------------------------------------------------------------
\section{Causal structure of a superluminal warp drive}\label{sect:causal}
%-------------------------------------------------------------------------------

We investigate the causal structure of a warp drive, following the method presented in \cite{causalstructure}, for spacetimes whose metrics can be written in Painlev\'e-Gullstrand coordinates. We begin from an eternal warp drive, moving at constant velocity, and then we study a dynamic situation in which a warp-drive bubble is accelerated.

%-------------------------------------------------------------------------------
\subsection{Eternal superluminal warp drive}
%-------------------------------------------------------------------------------
\label{subsec:etWD}

In 1+1 dimensions Alcubierre's metric Eq.~\eqref{eq:3Dalcubierre} reduces to
\begin{equation}\label{eq:alcubierre}
 ds^2=-c^2 dt^2+\left[dx-v(r)dt\right]^2~.
\end{equation}
 {Here $r$ is defined as the signed distance from the center of the bubble, $r\equiv x-v_0 t$. {Again we define $v(r)=v_0f(r)$ but $f$ is now taken to be defined also for negative values of $r$}. Its boundary conditions will 
be $f(0)=1$ and $f(r) \to 0$ for $r \to \pm \infty$.}
%providing of extending $f$ to negative value of its variable $r$.
 {For illustrative purposes, let us choose the following simple} bell-shaped function:
\begin{equation}\label{eq:f}
 f(r)=\frac{1}{\cosh\left(r/a\right)}
\end{equation}
and $v_0>c$, that is, our warp drive is superluminal. We want to stress that all the results of this paper do not depends on the particular choice of the function $f$. They are still valid providing that it satisfies the above conditions. We have chosen a particular form for it just for simplicity.
 {Now, by using $(t,r)$ coordinates the metric Eq.~\eqref{eq:alcubierre} reads}
%We change set of coordinates and write the metric (\ref{eq:alcubierre}) using $t$ and $r$. The center of the bubble has $r=0$ at any time. We obtain
\begin{equation}\label{eq:fluidmetric}
 ds^2=-c^2 dt^2+\left[dr-\bar{v}(r)dt\right]^2~,\qquad \bar{v}(r)=v(r)-v_0~.
\end{equation}
{Note that $\bar{v}<0$ because the warp drive is right going ($v>0$) but $v(r)\leq v_0$}.

 {By definition, at any time the center of the bubble is located at $r=0$}.
The causal structure of this metric can be analyzed following \cite{causalstructure}. Let us define the warp-drive Mach number $\alpha\equiv v_0/c$. The shift velocity becomes
\begin{equation}\label{eq:fluidvelocity}
 \bar{v}(r)=\alpha c\left[\frac{1}{\cosh\left(r/a\right)}-1\right]~.
\end{equation}
Two horizons appear when $\alpha>1$. Their positions are found by putting $\bar{v}$ equal to $-c$, \begin{equation}\label{eq:sonicpoints}
r_{1,2}
%_{\substack{1\\2}}
=\mp a \ln\left(\beta+\sqrt{\beta^2-1}\right),\qquad\beta\equiv\frac{\alpha}{\alpha-1}>1~.
\end{equation}
{Because of our simple profile choice, the two horizons are symmetrically located with respect to $r=0$. In more general situations $r_1$ and $r_2$ will be completely arbitrary satisfying only $r_1<r_2$.}

Right- and left-going null coordinates $u$ and $v$ can be defined as
\begin{align}
 du &\equiv dt - \frac{dr}{c+\bar{v}(r)}~, \label{eq:du}\\
 dw &\equiv dt + \frac{dr}{c-\bar{v}(r)}~. \label{eq:dw}
\end{align}
 {We note that the spacetime is divided into three distinct regions: $\rm I$ ($r<r_1$), $\rm II$ ($r_1<r<r_2$), and $\rm III$ ($r>r_2$). There are no $u$ rays connecting these regions, while $w$ rays cross all the regions; when $r$ approaches the horizons, $u$ diverges logarithmically. Integrating these equations we find
% \begin{widetext}
\begin{align}
\begin{split}
 u_{\rm I} &= t+\frac{r}{c\left(\alpha-1\right)}\\
	&-\frac{a\beta}{c\left(\alpha-1\right)\sqrt{\beta^2-1}}\ln\left[\frac{e^{-(r-r_1)/a}-1}{e^{-(r-r_2)/a}-1}\right]~,
 \end{split}\\
\begin{split}
 u_{\rm II} &= t+\frac{r}{c\left(\alpha-1\right)}\\
	&-\frac{a\beta}{c\left(\alpha-1\right)\sqrt{\beta^2-1}}\ln\left[\frac{1-e^{-(r-r_1)/a}}{e^{-(r-r_2)/a}-1}\right]~,
 \end{split}\\
\begin{split}
 u_{\rm III} &= t+\frac{r}{c\left(\alpha-1\right)}\\
	&-\frac{a\beta}{c\left(\alpha-1\right)\sqrt{\beta^2-1}}\ln\left[\frac{1-e^{-(r-r_1)/a}}{1-e^{-(r-r_2)/a}}\right]~,
 \end{split}\\
\begin{split}
 w &= t+\frac{r}{c\left(\alpha+1\right)}\\
	&+\frac{2a\gamma}{c\left(\alpha+1\right)\sqrt{1-\gamma^2}}\arctan\left(\frac{e^{r/a}-\gamma}{\sqrt{1-\gamma^2}}\right)~,
 \end{split}
\end{align}
% \end{widetext}
where $\beta$ is defined in Eq.~\eqref{eq:sonicpoints} and $\gamma\equiv\alpha/\left(\alpha+1\right)<1$.
In Fig.~\ref{fig:wd-uw} we plot the  lines of constant $u$ and $w$. 
\begin{figure}[t]
 \includegraphics[width=8.5cm]{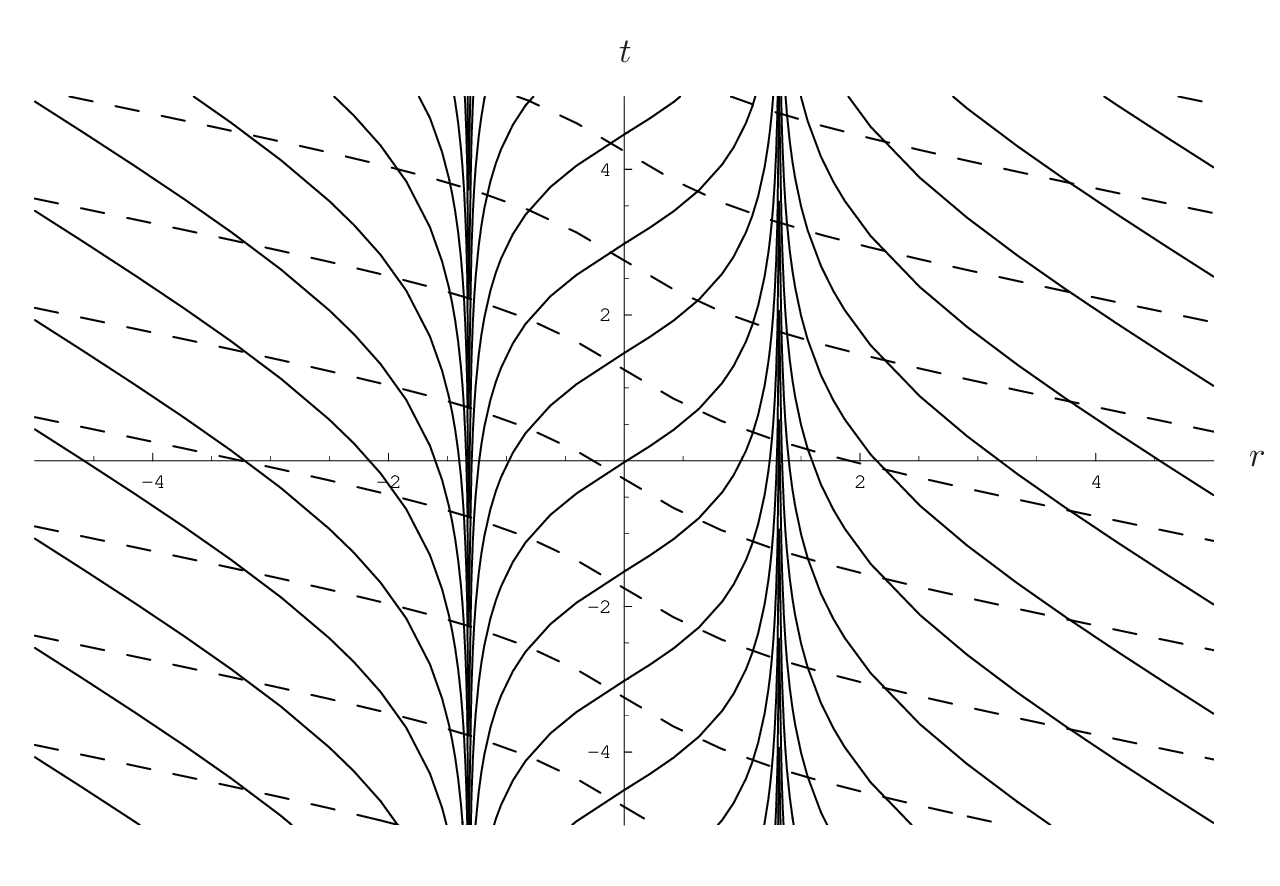}
 % wd-uw.eps: 1179666x1179666 pixel, 300dpi, 9987.84x9987.84 cm, bb=88 4 688 375
 \caption{Lines of constant $u$ (solid lines) and $w$ (dashed lines) for the Alcubierre warp drive, $a=c=1$, $\alpha=2$.}
 \label{fig:wd-uw}
\end{figure}
The horizon at $r=r_1$ corresponds to $u_{\rm I}\to +\infty$ and to $u_{\rm II}\to +\infty$, while the horizon
at $r=r_2$ corresponds to $u_{\rm II}\to -\infty$ and to $u_{\rm III}\to -\infty$.}

We define a signed surface gravity on our two horizons:
\begin{equation}
 \kappa_{1,2}
 %_{\substack{1\\2}}
 \equiv {\left.\frac{dv(r)}{dr}\right|}_{r=r_{1,2}
 %_{{\substack{1 \\ 2}}}
 }=\pm\frac{c\left(\alpha-1\right)\sqrt{\beta^2-1}}{a\beta}\equiv\pm \kappa ~.
\end{equation}
%Note that they have opposite values, such that the surface gravity $\kappa>0$ is the same for both the horizons.
{So doing, the surface gravity associated with the first horizon is positive $\kappa_1=\kappa>0$, while
the one associated with the second horizon is negative} $\kappa_2=-\kappa<0$.\footnote{In the definition of 
surface gravity, more correctly one should multiply the velocity derivative by $c$ to get an acceleration. However we use this slightly modified definition to avoid the appearance of too many constant $c$~factors in our formulas.}
As we noticed for the positions of $r_1$ and $r_2$, in a general situation in which $f$ is not symmetric, the two surface gravities may have different absolute values. However, the one associated with the first horizon (respectively, second horizon) will be always positive (respectively, negative). We will soon show that these two horizons represent a black and a white horizon, respectively.
Here on, we will consider the two surface gravities to have the same absolute value. Retaining different absolute values will not lead to more general results, but will just make the notation heavier.

\begin{figure}[t]
 \includegraphics[width=8.5cm]{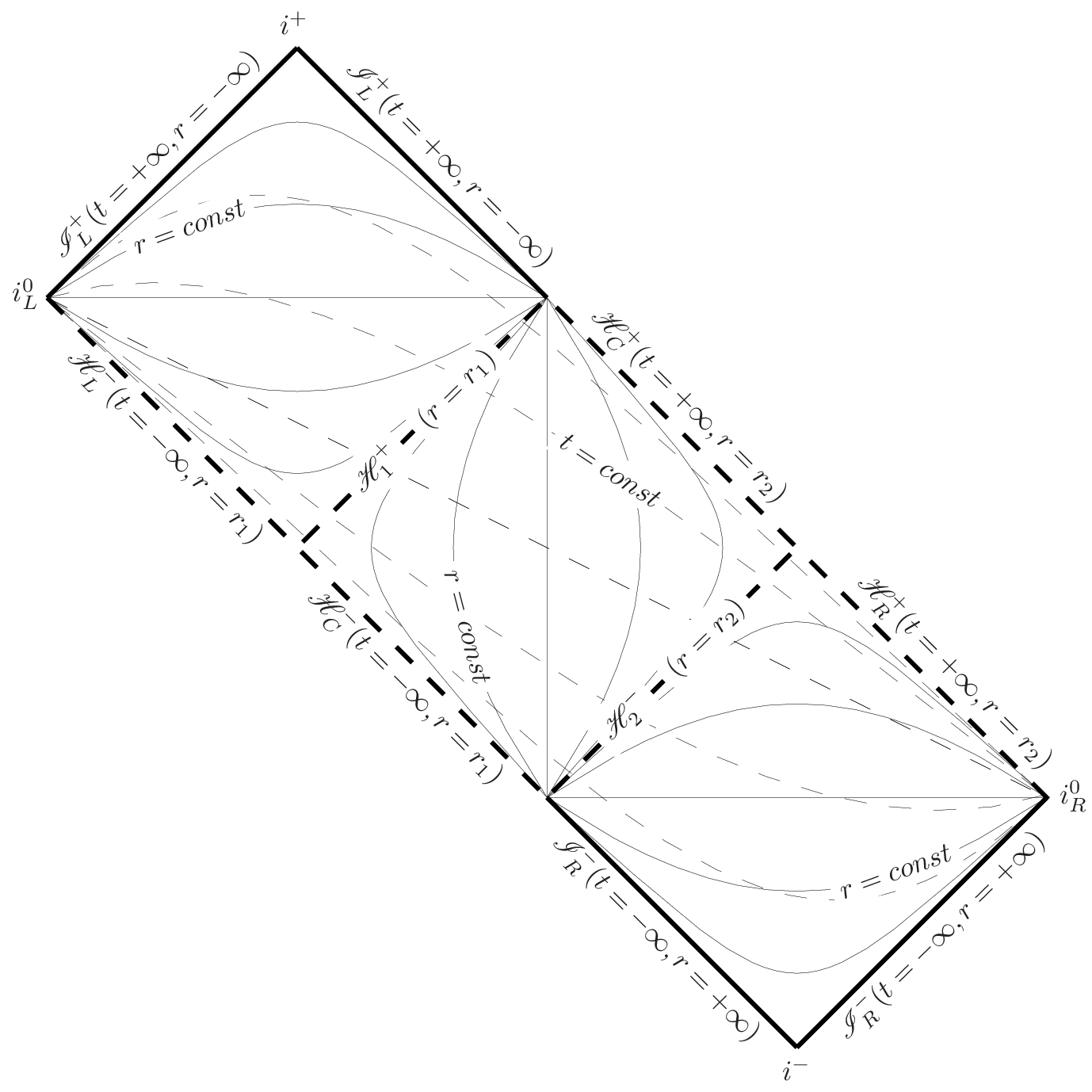}
 \caption{Penrose diagram of an eternal warp drive. Lines of constant $r$ (solid lines) and of constant $t$ (dashed lines). Future and past horizons at $r=r_1,r_2$ (heavy dashed lines). The geometry can be extended to the future of $\hor_C^+$ and $\hor_R^+$ and to the past of $\hor_C^-$ and $\hor_L^-$.}
 \label{fig:wd-pen}
\end{figure}

 {As in Sec. 6 of \cite{causalstructure}, one can find some transformation $U_i=U_i(u_i)$ that close to the horizons behaves as
\begin{eqnarray}
&&U_{\rm I}(u_{\rm I} \to +\infty) \simeq  U_{\rm I} + A_{\rm I} e^{-\kappa u_{\rm I}}~,
\\
&&U_{\rm II}(u_{\rm II} \to \pm \infty) \simeq U_{\rm II\pm} \mp A_{\rm II\pm} e^{\mp \kappa u_{\rm II}}~,
\\
&&U_{\rm III}(u_{\rm III} \to -\infty) \simeq U_{\rm III} - A_{\rm III} e^{\kappa u_{\rm III}}~,
\end{eqnarray}
where $U_{\rm I},U_{\rm II+},U_{\rm II-},\mbox{ and }U_{\rm III}$ are arbitrary constants and $A_{\rm I},A_{\rm II+},A_{\rm II-},\mbox{ and }A_{\rm III}$ are positive constants. It is possible to choose these transformations such that they match on the horizons, obtaining a global $U$ varying regularly from $+\infty$ to $-\infty$ as $r$ varies from $-\infty$ to $+\infty$.  {This matching will become natural when dealing with dynamical configurations in which the warp drive is created by accelerating the bubble from an initial zero velocity}. 
In order to do that we can choose $U_{\rm I}=U_{\rm II+}\equiv U_{\rm BH}$ and $U_{\rm II-}=U_{\rm III} \equiv U_{\rm WH}<U_{\rm BH}$.
%In order to make the overall transformation continuous and monotonically increasing the stationary region, that is inside the bubble, we must choose $U_{0+}>U_{0-}$. 
In this way the three transformation have the following domains:
\begin{align}
 u_{\rm I} \in (-\infty, +\infty)\quad &\Rightarrow\quad U_{\rm I} \in (+\infty, U_{\rm BH})~,\\
 u_{\rm II} \in (+\infty, -\infty)\quad &\Rightarrow\quad U_{\rm II} \in (U_{\rm BH}, U_{\rm WH})~,\\
 u_{\rm III} \in (-\infty, +\infty)\quad &\Rightarrow\quad U_{\rm III} \in (U_{\rm WH}, -\infty)~.
\end{align}

For instance, specific transformations having the required properties are the following:
\begin{align}
 &U_I=\frac{1}{2} + e^{-\kappa u_I}~,\\
 &U_{\rm II}=\frac{1}{2}\tanh\left(\frac{\kappa u_{\rm II}}{2}\right)~,\\
 &U_{\rm III}=-\frac{1}{2} - e^{\kappa u_{\rm III}}~.
\end{align}
%which can be written in the compact form
%\begin{multline}\label{eq:Utrans}
% U(u)=\theta\left(r_1 -r\right)\left(\frac{1}{2}+e^{-\kappa u}\right)\\
%	+\theta\left(r-r_1\right)\theta\left(r_2 -r\right)\frac{1}{2}\tanh\left(\frac{\kappa u}{2}\right)\\
%	-\theta\left(r-r_2\right)\left(-\frac{1}{2} - e^{\kappa u}\right),
%\end{multline}
%with $\theta$ the Heaviside unit step function. 
Now we can bring the right and left infinities to a finite distance by using a compactifying transformation like}
%Now we can compactify the spacetime with the following transformation
\begin{align}
 \mathcal{U}_{\rm I} &\equiv \arctan(U_{\rm I})~,\\
 \mathcal{U}_{\rm III} &\equiv \arctan(U_{\rm III})~,\\
 \mathcal{W} &\equiv \arctan(w)~.
\end{align}
 {The 
%and plot the 
Penrose diagram for this spacetime is plotted in Fig.~\ref{fig:wd-pen}. Notice that the diagram does not correspond
to a maximal analytical extension but to a particular patch of the total spacetime. The dashed lines signal the locations at which the geometry can be extended.}  
%In this diagram two horizons are clearly present  {displayed}. With respect to an observer inside the bubble at $r=r_2$ we have a past horizon \hm~and at $r=r_1$ we have a future horizon \hp. 
The two external regions $r>r_2$ and $r<r_1$ appear to an observer living inside the bubble as an eternal white hole and an eternal black hole, respectively. For this reason, we call \hp~and \hm, respectively, the \emph{black horizon} and the \emph{white horizon} of the warp drive. Notice however that for the inner observers in an eternal configuration, at $r_1$ there is also a white horizon [$\hor_C^-(t=-\infty, r=r_1)$ in the diagram] and at $r_2$ a black horizon [$\hor_C^+(t=+\infty, r=r_2)$ in the diagram]. The geometry can be extended through these two null lines. We do not picture the extended regions as there are ambiguities in the prescription of the matter distribution in those other ``universes."\footnote{In the analysis of the maximal extention of the Schwarzschild or Kerr spacetimes this problem is not present as one considers only vacuum solutions of the Einstein equations.}

%----------------------------------------------------------------------
\subsection{Dynamic superluminal warpdrive}
%---------------------------------------------------------------------- 
\label{subsec:dynWD}

What happens to the causal structure when we consider the creation of a superluminal warp drive starting from initially flat spacetime? For concreteness, we study a simple case in which we reach the final velocity $v=v_0$ at a finite time which we take to be $t=0$. We modify the metric Eq.~\eqref{eq:alcubierre} introducing a switching factor $\delta(t)$:
\begin{equation}\label{eq:dynamicalcubierre}
 ds^2=-c^2 dt^2+\left[dx-v(r,t)dt\right]^2~,
\end{equation}
where
\begin{equation}
v(r,t)=v_0\delta(t)f(r)~,
\end{equation}
with  $f(r)$ defined in Eq.~\eqref{eq:f} and
\begin{equation}
 \delta(t)\equiv
 \left\{
  \begin{aligned}
    &e^{t/\tau}\qquad &\text{if} \quad t<0~,  \\
    &1\qquad &\text{if}\quad t \geq 0~.
  \end{aligned}
 \right.
\end{equation}
Again we can change coordinates, keeping the center of the bubble at rest ($r=0$). This can be obtained by defining
\begin{equation}
 dr\equiv dx-v_0\delta(t) dt~.
\end{equation}
This is an exact differential form and can be integrated to get
\begin{equation}
 r=
 \left\{
  \begin{aligned}
    &x-v_0\tau\left[e^{t/\tau}-1\right]\qquad &\text{if} \quad t<0~,  \\
    &x-v_0 t\qquad &\text{if}\quad t\geq 0~.
  \end{aligned}
 \right.
\end{equation}
Again, the metric becomes
\begin{equation}\label{eq:dynamicfluidmetric}
\begin{split}
 &ds^2 =-c^2 dt^2+\left[dr-\hat{v}(r,t)dt\right]^2~,\\
 &\hat{v}(r,t) =v_0\delta(t)\left[f(r)-1\right]
\end{split}
\end{equation}
and the apparent horizons associated with the $t$~slicing are found by setting $\hat{v}=-c$. In this case a solution does not exist for any value of $t$, so that the apparent horizons are created at infinity at some finite $t_H$. We show this below. Let us write the equation for the apparent horizons in the following form:
\begin{equation}\label{eq:eqsonic}
 f(r)=1-\frac{c}{v_0\delta(t)}~.
\end{equation}
The function $f$ takes all the values between $0$ and $1$. In particular, $f(r)\to 0$ for $r\to\pm\infty$ and $f(0)=1$. Besides, the right-hand side of Eq.~\eqref{eq:eqsonic} is a monotonic function of $t$, such that, for $t\to-\infty$, $1-c/\left(v_0\delta(t)\right)\to-\infty$ and reaches the value $1-c/v_0>0$ for $t \geq 0$. As a consequence, there exists a time $t_H<0$ 
so that for $t>t_H$ there are always two solutions of Eq.~\eqref{eq:eqsonic}, corresponding to a black and a white horizon. These horizons are born at $t=t_H$ at spatial infinity and at $t=0$ they settle at two fixed positions $r_1$ and $r_2$. 

\begin{figure}[t]
\includegraphics[width=8.5cm]{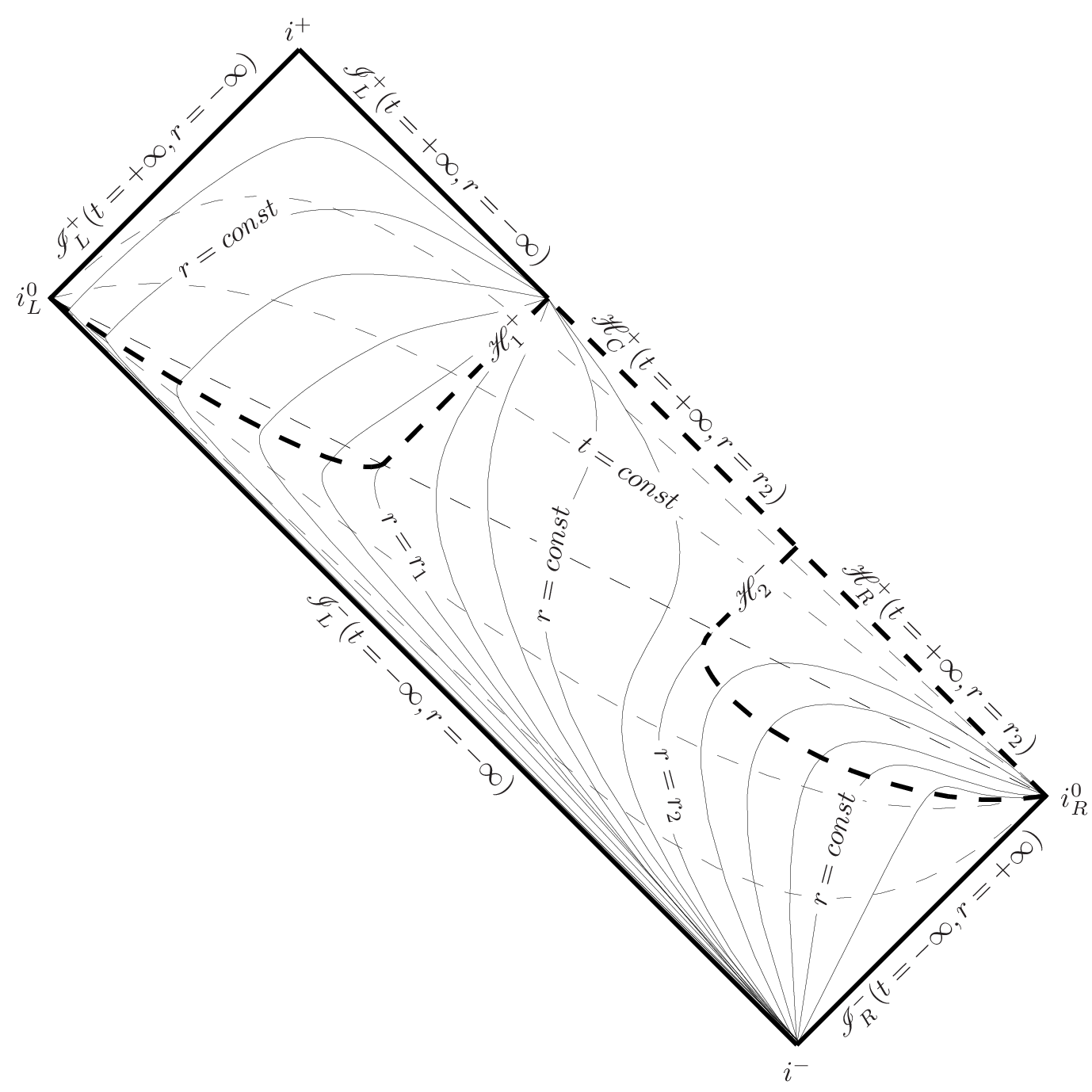}
 \caption{Penrose diagram of a dynamic warp drive. Lines of constant $r$ (solid lines) and of constant $t$ (dashed lines). The lines of constant $r$ become null at the apparent horizons (heavy dashed lines).}
 \label{fig:dynwd-pen}
\end{figure}

Keeping these points in mind, we are able to build the Penrose diagram for the dynamic warp drive~(Fig.~\ref{fig:dynwd-pen}). At early times the metric is approximately Minkowskian, because $\delta(t)\to0$ for $t\to-\infty$. 
Therefore, the causal structure is initially Minkowskian. Then, it progressively changes till $t=0$. At this time one has built a stationary warp drive, just as in the previous section.
% The causal structure is therefore Minkowskian and remains, so no horizons are present in the spacetime, i.e., till $t=t_H$. Then, a transition period takes place till $t=0$. At this time we have built a stationary warp drive, just like in the previous section.
After this time, the Penrose diagram looks exactly equal to that in Fig.~\ref{fig:wd-pen}. The final Penrose diagram is just obtained by gluing together the two behaviors. Again, we are not drawing an analytically extended diagram but only the relevant patch for the analysis that follow in this paper. {Given that a timelike observer can reach   \scribub~and $\hor_R^+$ in a finite proper time, the geometry can be extended in the future, beyond these lines. This is actually a subtle point. In fact,  \scribub~and $\hor_R^+$ (which are linked to the formation of a white horizon) are on the boundary of the Cauchy development of \scrim. In this sense, they are Cauchy horizons given that initial data are assigned only on \scrim. Hence, as noticed in Sec.~\ref{subsec:etWD}, an eventual extension would not be unique. In any case such an extension will not be relevant for what will follow, given that we shall limit ourselves to investigating the asymptotic behavior of the RSET associated with the onset of the superluminal warp drive.} 
%Beyond \scribub, the space time can be extended, so, if the larger manifold was well-behaved enough, one could find in this manifold a larger achronal set $\mathcal{S}\supset\scri_-$, so that \scribub~would not be a Cauchy horizon, and, according to Fulling-Sweeny-Wald theorem, the divergence should disappear. 
%However, as noticed in Sect.~\ref{subsec:etWD}, this extension is not unique, so we cannot reach a definitive answer on this regard.}

Let us highlight here that the dynamic way we have used here to create the warp drive is not the only possible way. In Appendix~\ref{app:kink}, we will also use a different interpolation between Minkowski and the warp drive in which the horizons appear also at finite time but at finite $r$ positions (similar to what happens when a homogeneous star collapses to form a black hole).

%-----------------------------------------------------------------
\section{Light-ray propagation}\label{sect:radiation}
%-----------------------------------------------------------------

The just discussed causal structure of the dynamical warp drive is naturally leading to the expectation that some sort of Hawking radiation will be produced in a superluminal warp drive (as well as some transient particle emission). It is well known (see, for example,~\cite{birreldavies}) that all the information about particle production is encoded in the way in which light rays propagate in a spacetime. That is, it is enough to know how light rays are bended in order to analyze the phenomenon of particle creation.
In the dynamical warp drive there is a single past null coordinate but three different future null coordinates associated with the final regions $\rm I$, $\rm II$ and $\rm III$, as described before. From here on, we will be dealing exclusively with the connection between the past null coordinate $U$ at $\scri_L^-$ and the future null coordinate $u_{\rm II}$ at \scribub~in the interior of the bubble. Therefore, we will use $u$ to denote $u_{\rm II}$ whenever this does not lead to confusion. As discussed in \cite{particlecreation} the relation $U=p(u)$ encodes all the relevant information about particle production.

We want to study the features due to the two main properties of a dynamical warp-drive geometry, i.e., the spacetime is Minkowskian at early times and it is a warp drive at late times. In particular, we are not interested in the transient features depending on how the transition between these two regimes is performed. Namely, we need only the behavior close to the horizons and at late times inside the whole bubble. It is clear from Fig.~\ref{fig:dynwd-pen} that, if one stays at constant $r$ inside the bubble and moves forward in time one crosses $u$ rays which pass closer and closer to the black horizon. Therefore, once we have determined the behavior of $p(u)$ close to the horizons,\footnote{Appendix~\ref{app:kink} gives a specific example for which this relation can be computed exactly in the whole spacetime.} we automatically also have the required information at late times in the whole bubble.

In general, the relation $U=p(u)$ is obtained by integrating the differential equation for the propagation of right-going light rays
\begin{equation}\label{eq:rays}
 \frac{dr}{dt}=c+\hat{v}(r,t)~.
\end{equation}
{Note that while in the previous section we considered a specific form of  $\hat{v}(r,t)$ [Eq.~\eqref{eq:dynamicfluidmetric}] in order to discuss the causal structure of the associated spacetime, here (and in what will follow) the discussion will hold for any $\hat{v}(r,t)$ that satisfies the requirements $\hat{v}(r,t)\to0$, for $t\to-\infty$  (sufficiently rapid for the spacetime to be asymptotically flat), and $\hat{v}(r, t)=\bar{v}$ after some finite time and within the warp-drive bubble.} 
Given the assumption that the spacetime settles down to a stationary warp-drive configuration, at late times the velocity profile will depend only on the $r$ coordinate. We can write, as in the stationary case Eq.~\eqref{eq:du}
\begin{equation}\label{eq:udef}
 du=dt-\frac{dr}{c+\bar{v}(r)}~.
\end{equation}
To find the required asymptotic relation one has to integrate this equation in the limit $r\to r_{1,2}$.
%{\substack{1 \\ 2}}$. 
There, the velocity can be expanded as
\begin{equation}
 \bar{v}=-c\pm \kappa \left(r-r_{1,2}
 %{\substack{1 \\ 2}}
 \right)+{\cal O}\left(\left(r-r_{1,2}
 %{\substack{1 \\ 2}}
 \right)^2\right)~.
\end{equation}
Thus, we obtain
\begin{equation}
 u\simeq t\mp\frac{1}{\kappa}\ln\left|r-r_{1,2}
 %{\substpeack{1 \\ 2}}
 \right|~.\label{eq:uwhapprox}
\end{equation}
On the other hand, the coordinate $U$, obtained by integrating Eq.~\eqref{eq:rays} at early times, reduces to the Minkowski null coordinate
\begin{equation}
 U(t\to -\infty)=t-\frac{r}{c}~,
\end{equation}
and is regular in the whole spacetime, in particular, on the horizons. For instance, on a fixed $t$~slice in the stationary region, we can write $U$ as a regular function of $r$
\begin{equation}\label{eq:Uanalitic}
 U_\pm=\mathcal{U}_\pm\left(r-r_{1,2}
% {\substack{1 \\ 2}}
 \right)~,
\end{equation}
where we denoted with $U_+$ (respectively, $U_-$) the specific form of $U$ close to the black (respectively, white) horizon, and $\mathcal{U}\pm$ are analytic functions. Inserting Eq.~\eqref{eq:uwhapprox} in the above expression, at the same fixed time,
\begin{equation}
 U_\pm=p(u\to\pm \infty)=\mathcal{P}_\pm(e^{\mp\kappa u})~,
\end{equation}
where $\mathcal{P}_\pm$ are again analytic functions. Note that the forms of these functions do not depend on the particular time slice chosen to perform the matching between Eqs.~\eqref{eq:uwhapprox} and \eqref{eq:Uanalitic}. In the proximity of the stationary horizons $u\to\pm\infty$, so $e^{\mp\kappa u}\to 0$ and the function $p$ can be expanded around the horizons. Up to the first order we get
\begin{equation}\label{eq:Uu}
 U=p(u \to \pm \infty)= U_{\substack{\rm BH \\ \rm WH}} \mp A_{\pm} e^{\mp\kappa u}+{\cal O}\left(e^{\mp2\kappa u}\right)~,
\end{equation}
where $A_{\pm}$ are positive constants.

This is indeed the asymptotic behavior one would expect in the presence of trapping horizons and, for the black hole case, it is the standard relation between $u$ and $U$. In fact, it leads to the conclusion that an observer at \scrip~will detect Hawking radiation with temperature $\kappa /2\pi$.
% While this results was obtained analytically thanks to the simplification allowed by the special velocity profile we adopted (with a kink), 
It is important to note that the result is completely general. The asymptotic behavior of $U=p(u)$ for large absolute values of $u$,
which is the only feature of $p(u)$ relevant for the analysis of this paper,
%  which is only relevant for the analysis of this paper,
does not depend on the specific velocity profile adopted. It is only necessary that it interpolates from Minkowski spacetime at early times to a stationary warp-drive geometry at late times.

%that all the following analytical computations will make use only of the above given asymptotic behavior of $U(u)$ for large absolute values of $u$. Therefore the qualitative features do not depend on the specific velocity profile adopted once the same asymptotic behaviour is recovered. Given that Eq.~\eqref{eq:Uu} is recovering the standard light rays ``peeling off"  in proximity of a trapping horizon we do expect that our result is generic. 

%For the white horizon, this gives rise to something similar to ingoing Hawking radiation in a white hole originated by a collapse \cite{hawkrad}. 
While Eq.~\eqref{eq:Uu} is exactly of the expected form, its implications for particle production are not as straightforward as in the black hole case. In fact, in the warp-drive geometry the late-time modes labeled by $u$ will not be standard plane waves in an asymptotically flat region of spacetime as they will be characterized by the strange form given in Eq.~\eqref{eq:uwhapprox}.
Of course, if the surface gravity $\kappa$ is large enough so that the typical wavelength of the emitted radiation is much smaller than the bubble size, then the plane-wave approximation is fine and in the center of the bubble, at late times, one will measure standard Hawking radiation at temperature $T$. Nonetheless, in the general case, even if the calculation for the Bogoliubov coefficient is the standard one \cite{birreldavies}, the particles created will not be standard plane waves. The physics associated with the particle production by the white horizon is even less clear. This is why, in the next section, we shall consider the behavior RSET to get more significant information.

\begin{figure}[tb]
 \includegraphics{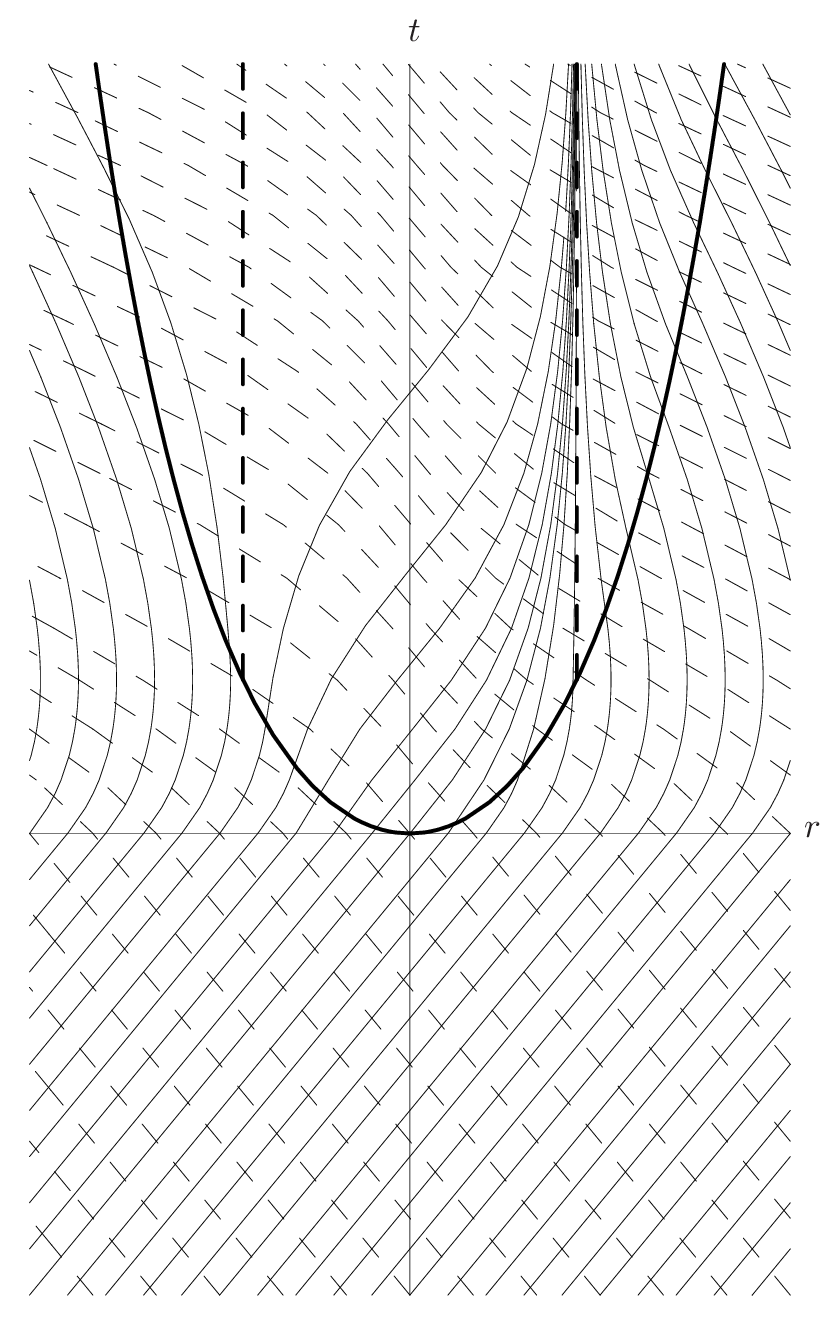}
 \caption{Light rays propagating rightward (\emph{solid lines}) and leftward (dashed lines)  in the plane $(t,r)$ in a warp-drive spacetime with velocity profile of Eq.~\eqref{eq:ppwarpvelocity}. The out region in which the geometry is a stationary warpdrive is at $r<\pm\text{arccosh}(t+1)$ (heavy solid lines). At $t<0$ the metric is Minkowskian. The horizons at $r_1$ and $r_2$ (heavy dashed lines) are formed at $T_H=1$. Please refer to Appendix~\ref{app:kink} for details.}
 \label{fig:lightrayssimp}
\end{figure}

In order to do so, we shall also need the relation between ingoing and outgoing left-going rays. In fact, these modes are excited too when the warp drive forms, even if we do not have a thermal particle production as for right-going modes. Left-going rays are the solution of the following differential equation [see Eq.~\eqref{eq:rays}]:
\begin{equation}\label{eq:raysleft}
 \frac{dr}{dt}=-c+\hat{v}(r,t)~.
\end{equation}
Looking at Fig.~\ref{fig:dynwd-pen}, we note that left-going rays do not see the horizons, that is, they cross them from $\scri^-_R$ to $\scri^+_L$. As a consequence, both the past and the future null coordinates $W$ and $w$ are defined at the asymptotic region outside the bubble. However, after the geometry inside the bubble has settled down to its final stationary form, we can define, just for convenience, a different coordinate $\tilde w$ inside the bubble, as in Eq.~\eqref{eq:dw}:
%However we do not have a natural definition of the coordinate $w$ for left-going rays as in Eq. (\ref{eq:ulimit}).  However, what we need is a null coordinate such that, for each ray, $w$ is an integration constant of Eq. (\ref{eq:raysleft}) beyond the kink, where $v$ depends only on $r$. From this point of view we see that Eqs. (\ref{eq:Ulimit}), (\ref{eq:ulimit}) and (\ref{eq:Wlimit}) determine the zero-point respectively of $U$, $u$, $W$ in a natural way. For $W$ this is not possible, so we decide arbitrarily its zero-point. We define
\begin{equation}\label{eq:wdef}
 d\tilde w=dt+\frac{dr}{c-\bar{v}(r)}~.
\end{equation}
Note that $w$ and $\tilde w$ may or may not coincide depending on how fast the metric in the external region settles down to its final stationary form (refer to Fig.~\ref{fig:lightrayssimp} and Appendix~\ref{app:kink} for an example in which they do not coincide).

$W$ is obtained in the {usual} way, by integrating Eq.~\eqref{eq:raysleft} at early times, when the spacetime is Minkowski
\begin{equation}
 W(t\to -\infty)=t+\frac{r}{c}~,
\end{equation}
The relation $W=q(\tilde{w})$ can be found explicitly for specific cases (see Appendix~\ref{app:kink}). The important point is that one can prove that this relation is always regular, so that it cannot give place to any phenomenon like Hawking radiation. Of course, one can choose a very unusual way to interpolate from Minkowski to the warp drive, such that a lot of particles are created in this sector, but this is not a general feature of dynamical warp drives. If we use a regular enough transition, only transient effects are present in this sector. In Appendix~\ref{app:kink}, we shall show that it is possible to find such a transition.

As an example, in Fig.~\ref{fig:lightrayssimp} we plot both right-going and left-going light rays propagating in the particular dynamic warp-drive spacetime analyzed in Appendix~\ref{app:kink}.

%---------------------------------------------------------------------
\section{Renormalized Stress-Energy Tensor}\label{sect:RSET}
%---------------------------------------------------------------------

For the calculation of the RSET inside the warp-drive bubble we can use the method used in \cite{stresstensor} for a 
collapsing configuration to form a black hole.
In null coordinates $U$ and $W$, affine on \scrim, the metric can be written as
\begin{equation}\label{eq:metricUW}
 ds^2=-C(U,W)dUdW~.
\end{equation}
As we described in the previous section, in the out region (the region in which the metric is stationary, i.e., $\bar{v}$ depends only on $r$) we can also use a different set of null coordinates $u, \tilde w$.
The coordinate $u$ is affine on $\hor_C^+$ and $\tilde w$ is the coordinate defined in~Eq.~\eqref{eq:wdef}. In these coordinates the metric is expressed as
\begin{equation}\label{eq:metricuw}
 ds^2=-\bar{C}(u,\tilde w) du d\tilde w~,
\end{equation}
{which implies}
\begin{equation}\label{eq:transfC}
 C(U,W) = \frac{\bar{C}(u,\tilde w)}{\dot{p}(u)\dot{q}(\tilde w)}~,
\end{equation}
and
\begin{equation}\label{eq:transdiff}
 U=p(u), \qquad W=q(\tilde w)~.
\end{equation}
These transformations are such that the $\bar{C}$ has precisely the form of the future stationary warp-drive geometry,
that is, it depends only on $r$ through $u,\tilde w$.

For concreteness let us refer to the RSET associated with having a single quantum massless scalar field living on the
spacetime. In this case the RSET components have the following form \cite{birreldavies}:
% {\tt (check the sign!!!)}
\begin{align}
 T_{UU} &= -\frac{1}{12\pi}C^{1/2}\partial_U^2 C^{-1/2}~,\label{eq:TUU}\\
 T_{WW} &= -\frac{1}{12\pi}C^{1/2}\partial_W^2 C^{-1/2}~,\label{eq:TWW}\\
 T_{UW} &= T_{WU}
 % =-\frac{1}{48\pi}\left(-\frac{C}{2}\right)R
 =\frac{1}{96\pi} C R~.\label{eq:TUWR}
\end{align}
Qualitatively, things would not change if there were other fields present in the theory. The only modification will be 
that the previous expressions will get multiplied by a specific numerical factor. It is clear that in the in region (where the spacetime is Minkowskian) the RSET is trivially zero.
%As we show in Appendix \ref{app:nullcoordinates}, to obtain this result is enough that the shift velocity does not depend on $r$.

Now, for a metric in the form of Eq.~\eqref{eq:metricUW}, the curvature can be calculated as in~\cite{birreldavies}:
\begin{multline}
 R=\Box\ln|C|={(-g)}^{-1/2}\partial_\mu\left[{(-g)}^{1/2}g^{\mu\nu}\partial_\nu\ln|C|\right]\\
	=-\frac{4}{C}\partial_U\partial_W\ln|C|~.
\end{multline}
Replacing this result in Eq.~\eqref{eq:TUWR} we obtain
\begin{equation}\label{eq:TUW}
 T_{UW}=T_{WU}=-\frac{1}{24\pi}\partial_U\partial_W\ln|C|~.
\end{equation}
Using transformations~\eqref{eq:transfC} and \eqref{eq:transdiff} in Eqs.~\eqref{eq:TUU}, \eqref{eq:TWW}, and \eqref{eq:TUW} we obtain
\begin{align}
 T_{UU} &= -\frac{1}{12\pi}\frac{1}{\dot{p}^2}\left[\bar{C}^{1/2}\partial_u^2 \bar{C}^{-1/2}-\dot{p}^{1/2}\partial_u^2 \dot{p}^{-1/2}\right]~,\label{eq:TUUbar}\\
 T_{WW} &= -\frac{1}{12\pi}\frac{1}{\dot{q}^2}\left[\bar{C}^{1/2}\partial_{\tilde{w}}^2 \bar{C}^{-1/2}-\dot{q}^{1/2}\partial_{\tilde{w}}^2 \dot{q}^{-1/2}\right]~,\label{eq:TWWbar}\\
 T_{UW} &= T_{WU}= -\frac{1}{24\pi}\frac{1}{\dot{p}\dot{q}}\partial_u\partial_{\tilde{w}}\ln|\bar{C}|~.\label{eq:TUWbar}
\end{align}

We can express the derivatives with respect to $u$ and $\tilde w$ in terms of derivatives with respect to $r$ and $t$. Note that the following expressions are obtained from Eqs.~\eqref{eq:udef} and \eqref{eq:wdef}, so they are valid only when the velocity profile depends only on the $r$ coordinate. In this fashion, we can study the stress-energy tensor at the end of the creation of the warp drive (or at the end of a collapse if we are studying a black hole).
\begin{multline}\label{eq:rtfromuw}
\begin{pmatrix}
\partial_r \\ 
\partial_t 
\end{pmatrix}
=
\begin{pmatrix}
 u_r & \tilde{w}_r\\
 u_t & \tilde{w}_t
\end{pmatrix}
\begin{pmatrix}
\partial_u \\
\partial_{\tilde{w}}
\end{pmatrix}
\\=
\begin{pmatrix}
  -1/(c+\bar{v}) & 1/(c-\bar{v})\\
  1 & 1
\end{pmatrix}
\begin{pmatrix}
\partial_u \\ 
\partial_{\tilde{w}}
\end{pmatrix}~.
\end{multline}
Inverting the derivative matrix we obtain the required result:
\begin{align}
 \partial_u &= -\frac{c^2-\bar{v}^2}{2c}\partial_r+\frac{c+\bar{v}}{2c}\partial_t~,\\
 \partial_{\tilde{w}} &= \frac{c^2-\bar{v}^2}{2c}\partial_r+\frac{c-\bar{v}}{2c}\partial_t~.
\end{align}
\comment{Of course one could calculate the derivatives in Eq.~\eqref{eq:rtfromuw} in a region where the metric is not time independent. In this case one cannot use anymore Eqs.~\eqref{eq:udef} and \eqref{eq:wdef}, but one can get the result passing through $U$ and $W$. Indeed, in these regions, $U$ and $W$ are defined in a very simple form as functions of $r$ and $t$ [see Eqs.~\eqref{eq:Utr} and \eqref{eq:Wtr}], so we can express $u$ and $\tilde{w}$ as functions of $U$ and $W$ and differentiate them. We get, when the velocity depends only on $t$:
\begin{multline}\label{eq:rtfromuwtime}
\begin{pmatrix}
\partial_r \\
\partial_t
\end{pmatrix}
=
\begin{pmatrix}
  u_r & \tilde{w}_r\\
 u_t & \tilde{w}_t
\end{pmatrix}
\begin{pmatrix}
\partial_u \\
\partial_{\tilde{w}}
\end{pmatrix}
=
 \begin{pmatrix}
 U_r/\dot{p} & W_r/\dot{q}\\
 U_t/\dot{p} & W_t/\dot{q}
\end{pmatrix}
\begin{pmatrix}
\partial_u \\ 
\partial_{\tilde{w}}
\end{pmatrix}
\\=
\begin{pmatrix}
  -1/\left(c\dot{p}\right)  & 1/\left(c\dot{q}\right)\\
  \left(1+\bar{v}/c\right)/\dot{p} & \left(1-\bar{v}/c\right)/\dot{q}
\end{pmatrix}
\begin{pmatrix}
\partial_u \\ 
\partial_{\tilde{w}}
\end{pmatrix}~.
\end{multline}
}
Please refer to Appendix~\ref{app:nullcoordinates} for more details.

We are interested in calculating the RSET inside the bubble when the two horizons have been formed and when the configuration has settled down to a stationary warp drive. Since in this region the velocity depends only on $r$, we can replace the derivatives in Eqs.~\eqref{eq:TUUbar}, \eqref{eq:TWWbar} and \eqref{eq:TUWbar} with
\begin{align}
 \partial_u &\rightarrow -\frac{1-\bar{v}^2}{2}\partial_r = -\frac{\bar{C}}{2}\partial_r~,\\
 \partial_{\tilde{w}} &\rightarrow  \frac{1-\bar{v}^2}{2}\partial_r = \frac{\bar{C}}{2}\partial_r~.
\end{align}
Here, we have put $c=1$, for simplicity, and we have used Eq.~\eqref{eq:barC}, $\bar{C}=1-\bar{v}^2$.
%and we have written $v$ instead of $\bar{v}$. In the last equality in both lines we used Eq. (\ref{eq:barC}) $\bar{C}=1-v^2$. 
Moreover we indicate with $'$ the differentiation with respect to $r$. After some calculations we obtain
\begin{gather}
 \bar{C}^{1/2}\partial_u^2 \bar{C}^{-1/2} =\bar{C}^{1/2}\partial_{\tilde{w}}^2 \bar{C}^{-1/2}=\frac{1}{16}\left[{\left(\bar{C}'\right)}^2-2\bar{C}\bar{C}''\right]~,\\
 \dot{p}^{1/2}\partial_u^2 \dot{p}^{-1/2} =\frac{1}{4\dot{p}^2}\left[3\ddot{p}^2-2\dot{p}\,\dddot{p}\right]~,\\
 \dot{q}^{1/2}\partial_{\tilde{w}}^2 \dot{q}^{-1/2} =\frac{1}{4\dot{q}^2}\left[3\ddot{q}^2-2\dot{q}\,\dddot{q}\right]~,\\
 \partial_u\partial_{\tilde{w}}\ln|\bar{C}| =-\frac{1}{4}\bar{C}\ddot{\bar{C}}~.
\end{gather}
Using again $\bar{C}=1-\bar{v}^2$ we get the final result:
\begin{align}
 T_{UU} &= -\frac{1}{48\pi}\frac{1}{\dot{p}^2}\left[\bar{v}'\,^2+\left(1-\bar{v}^2\right)\bar{v}\bar{v}''-\frac{3\ddot{p}^2-2\dot{p}\,\dddot{p}}{\dot{p}^2}\right]~,\\
 T_{WW} &= -\frac{1}{48\pi}\frac{1}{\dot{q}^2}\left[\bar{v}'\,^2+\left(1-\bar{v}^2\right)\bar{v}\bar{v}''-\frac{3\ddot{q}^2-2\dot{q}\,\dddot{q}}{\dot{q}^2}\right]~,\\
 T_{UW} &= T_{WU}=-\frac{1}{48\pi}\frac{1}{\dot{p}\dot{q}}\left(1-\bar{v}^2\right)\left[\bar{v}'\,^2+\bar{v}\bar{v}''\right]~.
\end{align}
One can check that these quantities do not diverge at the horizons, just like in \cite{stresstensor}. However we want to look at the energy density inside the bubble and try to understand whether it remains small or not as time increases.
\comment{
We need to write the stress-energy tensor components in the $(t,r)$ frame instead of the $(U,W)$ frame.
Thus we use Eq.~\eqref{eq:UWderivspq} to get
\begin{align}
\begin{split}\label{eq:Ttt}
 T_{tt}&=U_t^2 T_{UU}+2 U_t W_t T_{UW} +W_t^2 T_{WW} =\dot{p}^2 T_{UU}+2\dot{p}\dot{q}T_{UW}+\dot{q}^2 T_{WW}\\
&=-\frac{1}{48\pi}\left[2{\left(2-\bar{v}^2\right)}\bar{v}'\,^2+4\left(1-\bar{v}^2\right)\bar{v}\bar{v}''-f(u)-g(\tilde{w})\right],
\end{split}\\
\begin{split}\label{eq:Trr}
 T_{rr}&=U_r^2 T_{UU}+2 U_r W_r T_{UW} +W_r^2 T_{WW} =\frac{\dot{p}^2}{{\left(1+\bar{v}\right)}^2} T_{UU}-2\frac{\dot{p}\dot{q}}{1-\bar{v}^2}T_{UW}+\frac{\dot{q}^2}{{\left(1-\bar{v}\right)}^2} T_{WW}\\
&=-\frac{1}{48\pi}\left[\frac{2\bar{v}^2\left(3-\bar{v}^2\right)}{{\left(1-\bar{v}^2\right)}^2}\bar{v}'\,^2+\frac{4\bar{v}^3}{1-\bar{v}^2}\bar{v}''-\frac{f(u)}{{\left(1+\bar{v}\right)}^2}-\frac{g(\tilde{w})}{{\left(1-\bar{v}\right)}^2}\right],
\end{split}\\
\begin{split}\label{eq:Ttr}
  T_{tr}&=U_t U_r T_{UU}+ \left(U_t W_r + W_t U_r \right)T_{UW} +W_t W_r T_{WW} 
 =-\frac{\dot{p}^2}{1+\bar{v}} T_{UU}+2\frac{\bar{v}\dot{p}\dot{q}}{1-\bar{v}^2}T_{UW}+\frac{\dot{q}^2}{1-\bar{v}} T_{WW}\\
 &=-\frac{1}{48\pi}\left[\frac{2\bar{v}\left(2-\bar{v}^2\right)}{1-\bar{v}^2}\bar{v}'\,^2+4\bar{v}^2\bar{v}''+\frac{f(u)}{1+\bar{v}}-\frac{g(\tilde{w})}{1-\bar{v}}\right]~,
\end{split}
\end{align}
}

In particular, it is interesting to look at the energy measured by a set of free-falling observers, whose four velocity is simply $u_{c}^\mu=(1,\bar{v})$ in $(t,r)$ components.
\begin{widetext}
These observers measure an energy density $\rho$:
\begin{multline}
 \rho=T_{\mu\nu}u_c^\mu u_c^\nu=T_{tt}+2\bar{v}T_{tr}+\bar{v}^2T_{rr} =U_t^2 T_{UU}+2 U_t W_t T_{UW} +W_t^2 T_{WW}\\
 +2\bar{v}\left[U_t U_r T_{UU}+\left(U_t W_r + W_t U_r \right)T_{UW} +W_t W_r T_{WW}\right]+\bar{v}^2\left[U_r^2 T_{UU}+2 U_r W_r T_{UW} +W_r^2 T_{WW}\right]\\
 =\left(U_t+\bar{v} U_r\right)^2T_{UU}+2\left(U_t+\bar{v} U_r\right)\left(W_t+\bar{v} W_r\right)T_{UW}+\left(W_t+\bar{v} W_r\right)^2T_{WW}\\
 =\frac{\dot{p}^2}{\left(1+\bar{v}\right)^2}T_{UU}+2\frac{\dot{p}\dot{q}}{1-\bar{v}^2}T_{UW}+\frac{\dot{q}^2}{\left(1-\bar{v}\right)^2}T_{WW}\\
=-\frac{1}{48\pi}\left[\frac{2\left(\bar{v}^4-\bar{v}^2+2\right)}{\left(1-\bar{v}^2\right)^2}\bar{v}'\,^2+\frac{4\bar{v}}{1-\bar{v}^2}\bar{v}''-\frac{f(u)}{\left(1+\bar{v}\right)^2}-\frac{g(\tilde{w})}{\left(1-\bar{v}\right)^2} \right]~,
\end{multline}
\end{widetext}
where we have defined
\begin{align}
 f(u)&\equiv\frac{3\ddot{p}^2(u)-2\dot{p}(u)\,\dddot{p}(u)}{\dot{p}^2(u)}~,\label{eq:fu}\\
 g(\tilde{w})&\equiv\frac{3\ddot{q}^2(\tilde{w})-2\dot{q}(\tilde{w})\,\dddot{q}(\tilde{w})}{\dot{q}^2(\tilde{w})}~.
\end{align}

%Note that this set of observers are appropriate for our purposes. 
We want to study what happens to a spaceship placed at rest in the center of the bubble to investigate whether a warp drive can be used as a transportation device. Moreover we want to check whether the components of the RSET in a regular coordinate system are regular at the horizons. Freely falling observers are at rest in the center of the bubble but left
moving otherwise.

Looking at the above expressions for any component of the RSET or at the energy density $\rho$, we see that they can be split as a sum of three terms, one purely static, depending only on the $r$ coordinate through the shift velocity $\bar{v}$, and two dynamic pieces depending also on the $u$ and $\tilde{w}$ coordinates, respectively. They correspond to energy traveFling on right-going and left-going light rays, respectively, eventually red/blue-shifted by a term depending on $r$.
Being interested in studying the energy density measured by free-falling observers, we write
\begin{equation}
 \rho=\rho_{\rm st}+\rho_{{\rm dyn}-u}+\rho_{{\rm dyn}-\tilde{w}}~,\\
\end{equation}
where we defined static terms (labeled by subscript $\rm st$) and dynamic terms (labeled by subscript $\rm dyn$) for each component. Right-going and left-going terms are respectively labeled by $u$ and $\tilde{w}$.
\begin{gather}
 \rho_{\rm st}\equiv-\frac{1}{24\pi}\left[\frac{\left(\bar{v}^4-\bar{v}^2+2\right)}{\left(1-\bar{v}^2\right)^2}\bar{v}'\,^2+\frac{2\bar{v}}{1-\bar{v}^2}\bar{v}''\right]~,\label{eq:rhocs}\\
 \rho_{{\rm dyn}-u}\equiv\frac{1}{48\pi}\frac{f(u)}{\left(1+\bar{v}\right)^2}~,\label{eq:rhodu}\\
 \rho_{{\rm dyn}-\tilde{w}}\equiv\frac{1}{48\pi}\frac{g(\tilde{w})}{\left(1-\bar{v}\right)^2}~.\label{eq:rhodw}
\end{gather}
Let us start with $\rho_{{\rm dyn}-\tilde{w}}$. Its denominator is bounded for each $r$ because the shift velocity is negative, and its numerator is vanishing with time. It is easy to convince oneself that all the contributions to the RSET coming from the $\tilde{w}$ sector are not universal but depend exclusively on the specific interpolation between the early Minkowski spacetime and the final warp-drive spacetime. It is always possible to choose an interpolation so that all these contributions vanish at late times (see Appendix~\ref{app:kink}).

\comment{
We see from Fig.~\ref{fig:lightrayssimp} that, as time grows, points inside the bubble are reached by $\tilde{w}$ rays that intersect the kink later and later. Using Eqs.~\eqref{eq:wdef} and \eqref{eq:Wtr} we can estimate $\dot{q}$, calculating the derivatives of $\tilde{w}$ and $W$ with respect to $t$ on the kink, at crossing time $t_c$:
\begin{multline}
 \dot{q}=\frac{dW}{d\tilde{w}}=\left.\frac{dW}{dt}\right|_{r=\xi(t_c)}\left(\left.\frac{d\tilde{w}}{dt}\right|_{r=\xi(t_c)}\right)^{-1}\\
	=\left(1+\frac{\dot\xi(t_c)}{c}-\frac{\bar{v}(\xi(t_c))}{c}\right)\left(1+\frac{\dot\xi(t_c)}{c-\bar{v}(\xi(t_c))}\right)^{-1}\\
 =1-\frac{\bar{v}(\xi(t_c))}{c}\to 1-\frac{\bar{v}(+\infty)}{c},
\end{multline}
This is a constant value greater than $0$.
% Moreover $\dot{q}$ tends exponentially fast to this value, given a shift-velocity profile as in Eq. (\ref{eq:fluidvelocity}).
As a consequence, all the derivatives of $\dot{q}$ go to zero as time grows and $\rho_{{\rm dyn}-\tilde{w}}$ must go to zero in the same way. In conclusion, the dynamic term originating by the distorsion of left-going rays can be different from zero when the horizon is created. However this is only a transient term that is brought toward $\scri_L^+$, i.e., outside the bubble. That is, some radiation travelling leftward is produced only at the creation of the warp drive but there is not any phenomenon like Hawking radiation originated in this way. 
}
From now on, we neglect the dynamic term $\rho_{{\rm  dyn}-\tilde{w}}$ and study
\begin{equation}
 \rho\simeq\rho_{\rm  st}+\rho_{{\rm dyn}-u}~.\\
\end{equation}

%In order to estimate $\rho$ we use the asymptotic behavior of $f(u)$ given in Appendix \ref{app:fp}, Eqs. (\ref{eq:approxfu}) and (\ref{eq:approxf}).

\subsection{RSET at the center of the warp-drive bubble}
We shall now study the behavior of the RSET in the center of the bubble at late times. In this point $\bar{v}(r=0)=\bar{v}'(r=0)=0$ and the static term $\rho_{\rm st}$ vanishes.
By integrating Eq.~\eqref{eq:udef} in the stationary region, one can show that $u(t,r)$ is linear in $t$ so that, for fixed $r$, it will acquire arbitrarily large positive values (see Fig.~\ref{fig:dynwd-pen}).
% From Fig. \ref{fig:dynwd-pen}, we notice that for fixed $r$, if we wait enough time, $u(r,t)$ will assume arbitrary large positive values. This fact can easily check by integrating Eq.~(\ref{eq:udef}) and then fixing $r$ in the obtained expression, such that $u$ is a linear function of $t$.
% In this point $\bar{v}(r=0)=\bar{v}'(r=0)=0$ and the static term $\rho_{\rm st}$ vanishes. 
The dynamic term in Eq.~\eqref{eq:rhodu} can be evaluated by using a late-time expansion of $p(u)$ [see Eqs.~\eqref{eq:Uu} and~\eqref{eq:Uulong}], then plugged in Eq.~\eqref{eq:fu}.
This is derived in Appendix~\ref{app:fp} and it is equal to
% (\ref{eq:approxfu}) for $u_+\to+\infty$. We obtain
\begin{multline}
f(u)=\kappa^2\left\{1+\left[3{\left(\frac{A_{2+}}{A_{1+}}\right)}^2-2\frac{A_{3+}}{A_{1+}}\right]e^{- 2\kappa u}\right.\\
	\left.+{\cal O}\left(e^{- 3\kappa u}\right) \right\}~.
\end{multline}
With this expansion it is easy to see that for $u\to+\infty$
\begin{equation}
 \rho \simeq \frac{\kappa^2}{48\pi}~.
 \label{en-den}
\end{equation}
This result may be easily understood in the following way. The surface gravity of the black horizons is the velocity derivative evaluated at this horizon $\kappa=(d\bar{v}/dr)_{r=r_1}$. Moreover, the energy density of a scalar field at some finite temperature $T$ in $1+1$ 
dimensions is simply
%can be calculated as follows:
\begin{equation}\label{eq:scalarfield}
 \rho_{\rm  T}=\int\frac{\omega}{\left(e^{\omega/T}+1\right)}\frac{dk}{2\pi}
 %=\frac{1}{\pi}\int_0^{+\infty}\frac{\omega}{e^{\omega/T}+1}d\omega\\
	%=\frac{T^2}{\pi}\int_0^{+\infty}\frac{x}{e^x+1}dx
	=\frac{\pi}{12}T^2~.
\end{equation}
Defining the Hawking temperature in the usual way as $T_H\equiv \kappa/2\pi$, it is easy to see that Eq.~\eqref{en-den} can be rewritten exactly as $\rho=\left(\pi /12\right){T_H}^2$. 
Thus, the observer inside the warp-drive bubble will indeed observe thermal radiation at the temperature $T_H$.

%--------------------------------------------------------
\subsection{RSET at the bubble horizons}
\label{subsect:RSETH2}
%--------------------------------------------------------

Let us move now to study $\rho$ close to the horizons. Note that both $\rho_{\rm st}$ and $\rho_{{\rm dyn}-u}$ are divergent at the horizons ($r\to r_{1,2}$)
% _{{\substack{1 \\ 2}}}$)
because of the $(1+\bar{v})$ factors in the denominator. Just like in \cite{stresstensor}, for a black hole, these divergences exactly cancel each other, but something different happens at the black and white horizon. Expanding the velocity up to the second order one gets
\begin{equation}\label{eq:vseries}
 \bar{v}_\pm(r)=-1\pm\kappa\left(r-r_{1,2}
 %{\substack{1 \\ 2}}
 \right)+\frac{1}{2}\sigma \left(r-r_{1,2}
 %{\substack{1 \\ 2}}
 \right)^2 + {\cal O}\left(\left(r-r_{1,2}
 %{\substack{1 \\ 2}}
 \right)^3\right)~.
\end{equation}
where $\sigma$ is a constant.

Close to the horizons the static term of Eq.~\eqref{eq:rhocs} then becomes
\begin{equation}
 \rho_{\rm st}\left(r \simeq r_{1,2}
 %{\substack{1 \\ 2}}
 \right)
 %-\frac{1}{24\pi}\left[\frac{\left(\bar{v}^4-\bar{v}^2+2\right)}{\left(1-\bar{v}^2\right)^2}\bar{v}'\,^2+\frac{2\bar{v}}{1-\bar{v}^2}\bar{v}''\right]\\
	=-\frac{1}{48\pi}\left[\frac{1}{\left(r-r_{1,2}
%{\substack{1 \\ 2}}
\right)^2}\mp\frac{\sigma}{\kappa \left(r-r_{1,2}
%{\substack{1 \\ 2}}
\right)}\right]+{\cal O}(1)~.
\end{equation}
Similarly one can expand the dynamic term $\rho_{{\rm dyn}-u}$ to the same order. 
This involves determining the function $f(u)$ in the proximity of the horizons. Indeed it is not difficult to show that (see Appendix~\ref{app:fp})
\begin{widetext}
\begin{equation}
\lim_{r\to r_{\pm} }f(u) = \kappa^2\left\{1+\left[3{\left(\frac{A_{2\pm}}{A_{1\pm}}\right)}^2-2\frac{A_{3\pm}}{A_{1\pm}}\right]e^{\mp 2\kappa t}\left(r-r_{1,2}
 %{\substack{1 \\ 2}}
\right)^2+{\cal O}\left(\left(r-r_{1,2}
 %{\substack{1 \\ 2}}
\right)^3\right) \right\}~.
\end{equation}
\end{widetext}
Inserting this form in Eq.~\eqref{eq:rhodu} we obtain:
\begin{equation}
 \rho_{{\rm dyn}-u}\left(r \simeq r_{1,2}
 %{\substack{1 \\ 2}}
\right)
 %=\frac{1}{48\pi}\frac{f(u)}{\left(1+\bar{v}\right)^2}\\
	= \frac{1}{48\pi}\left[\frac{1}{\left(r-r_{1,2}
 %{\substack{1 \\ 2}}
\right)^2}\mp\frac{\sigma}{\kappa\left(r-r_{1,2}
 %{\substack{1 \\ 2}}
\right)}\right]+{\cal O}(1)~,
\end{equation}
It is now clear that the total $\rho$ is ${\cal O}(1)$ on the horizon and does not diverge (as expected from the Fulling-Sweeny-Wald theorem \cite{fsw}).

However, let us look to the subleading terms
%is evident at the white horizon. In fact, the subleading terms are $O(1)$ at the formation of the horizon but are at the white horizon exponentially growing time.
%More explicitly one finds
 %Note that the static term does not depend on time. Using fully Eq. (\ref{eq:approxf}), we see that, after having cancelled the divergent piece in the coordinate $x$ and looking at the behavior at late times:
\begin{multline}
 \rho\left(r \simeq r_{1,2}
 %{\substack{1 \\ 2}}
\right)=\frac{1}{48\pi}\left[3{\left(\frac{A_{2\pm}}{A_{1\pm}}\right)}^2-2\frac{A_{3\pm}}{A_{1\pm}}\right]e^{\mp2\kappa t}\\
	+C_\pm+{\cal O}\left(r-r_{1,2}
 %{\substack{1 \\ 2}}
\right)~,
\label{eq:sublead}
\end{multline}
where $C_\pm$ are constants. We can easily see that the behavior close to the black horizon is completely different from that close to the white horizon. In the former case the energy density as seen by a free-falling observer is damped exponentially with time.
%~\footnote{Note however, that in analogy to the conclusions drawn in \cite{stresstensor}, also in this case a slow approach to the horizon formation might lead to potentially large values of the RSET and hence large back reaction which would oppose to the creation of the horizons.} 
In the white horizon, however, this energy density grows exponentially with time. {This means that, moving along $\hor_2^-$, $\rho$ is large and negative diverging while approaching the crossing point between $\hor_2^-$ and~\scribub}.
% According to the results obtained by \cite{hawkrad}, this is due to the transient ingoing radiation produced by the white horizon. However radiation cannot cross the horizon from the bubble to the exterior region, being this horizon white. Thus radiation must pack into it.

This asymptotic divergence is physical and not a matter of selection of coordinates. In a very short time after the white horizon is formed (of the order of $1/\kappa$), the backreaction of the RSET in this region of spacetime is no longer negligible {but rather very strong}. 
%This growing behavior can be appreciated in Fig. \ref{fig:rhocwarp}, where $\rho$ is plotted, for the warpdrive model presented in Appendix~\ref{app:kink}.
%\begin{figure}
% \psfragax
%  \includegraphics[width=0.47\textwidth]{../figures/rhorot0.5.eps} 
%% \includegraphics[width=0.5\textwidth]{../figures/rhocont0.5.eps}
% \caption{Plot of $\rho$ inside the bubble ($r_1<r<r_2$) after the horizons have been formed $t>t_H=1$. $\alpha=2$, $a=1$ and $\xi(t)=\text{arccosh}(t+1)$ (see Appendix~\ref{app:kink} for details). The plot range is $0<t<3$ (\emph{vertical axis}), $r_1<r< r_2$ (\emph{horizontal axis}). {Note the two different divergences. In the former one, exactly at $r=r_2$, ($\hor_2^-$ in the Penrose diagram, of Fig.~\ref{fig:dynwd-pen}) and for $t\to+\infty$ (approaching the crossing point of $\hor_2^-$ and \scribub), $\rho\to-\infty$. In the latter one, $\rho\to+\infty$ for  $t\to+\infty$ and $r\to r_2$ (approaching any point of \scribub. Note that, in a $t,r$ plot,  the couple $(t=+\infty,r=r_2)$ represents a point, while in the conformal diagram is the whole segment \scribub. The two different divergences appear in the $t,r$ plot when one approaches the point $(+\infty,r_2)$ along different curves.}}
% \label{fig:rhocwarp}
%\end{figure}

%-------------------------------------------------------------------------
\subsection{RSET at late times approaching \scribub }
\label{subsect:RSETHc}
%-------------------------------------------------------------------------

In Sec.~\ref{subsect:RSETH2} we studied the RSET on the black and white horizons. However, it is interesting to analyze better its behavior as it approaches the Cauchy horizon \scribub. As one can see from Fig.~\ref{fig:lightrayssimp}, every $u$ ray reaches values of $r$ very close to $r_2$, at  sufficiently late times. This means that, even for large positive values of $u$, some time exists after which the approximate behavior of all the $u$ rays is well described by
\begin{equation}
 u\simeq t+\frac{1}{\kappa}\ln{\left(r_2-r\right)}~,
\end{equation}
so that
\begin{equation}
 r_2-r\propto e^{-\kappa t}~.
\end{equation}
Keeping in mind this point, let us study the term $\rho_{{\rm dyn}-u}$, at fixed $u$, when time increases. There is a time at which the $u$ ray will be close enough to $r_2$, such that the denominator in Eq.~\eqref{eq:rhodu} can be approximated by
\begin{equation}
 {\left(1+\bar{v}\right)}^2\simeq \kappa^2 {\left(r_2-r\right)}^2~,
\end{equation}
which becomes, thanks to the previous result,
\begin{equation}
 {\left(1+\bar{v}\right)}^2\propto e^{-2\kappa t}~,
\end{equation}
such that, for $t\to+\infty$, $\rho$ diverges along every single $u$ ray as $e^{+2\kappa t}$, because of the blueshift factor $(1+{\bar{v}})^2$. 
Hence, $\rho$ will diverge on the whole line \scribub, which is a Cauchy horizon for the geometry (so this result does not contradict the Fulling-Sweeny-Wald theorem~\cite{fsw}).  We then deduce that the warp-drive spacetime is again likely to become unstable in a very short time.

As a closing remark, it is perhaps important to stress that this divergence of the RSET on the Cauchy horizon is of different nature with respect to the one found in Sec.~\ref{subsect:RSETH2}.
In fact, the divergence of Sec.~\ref{subsect:RSETH2} is intrinsically due to the inevitable transient disturbances produced by the formation of the horizon.
In this sense it is a new and very effective instability present every time a white horizon is formed in some dynamical way.
On the contrary, the just found divergence on \scribub~can be seen as due to the well-known infinite blueshift suffered by light rays as they approach a Cauchy horizon, in this specific case as due to the accumulation of Hawking radiation produced by the black horizon.
In this sense it is not very different from the often claimed instability of inner horizons in Kerr-Newman black holes~\cite{Simpson:1973ua,poissonisrael,markovicpoisson}.

%Note that the divergence found in the previous subsection is quite different from this one.
%%In that case it 
%{In the former one, the divergence} was due only to particle production at the white horizon, that is, it would be there even if only a white hole was present.
%{The divergence $\rho \to -\infty$ appears exactly on $\hor_2^-$ and for $t \to +\infty$.}
%This latter divergence, instead, is due to the whole warpdrive geometry, in the sense that every form of energy (both transient terms appearing in the transition region and Hawking radiation created at the black horizon) travelling rightward will cause a divergence of~$\rho$.
%{In this divergence $\rho\to +\infty$ as one approaches any point of \scribub.}
%%{In Fig.~\ref{fig:rhocwarp}~we can appreciate these two different divergences. 
%%Therefore, the backreaction of the RSET will be very strong, making the warp-drive spacetime to become unstable.

Of course the divergence and the appearance of the horizon \scribub~would be avoided if the superluminal warp drive were sustained just for a finite amount of time. In that case, no Cauchy horizons would arise and no actual infinity would be reached by the RSET.  However, the latter would still become huge in a very short time, increasing exponentially on a time scale $1/\kappa\approx \Delta/c$, where $\Delta$ is the thickness of the warp-drive bubble. Note that, in order to get a time scale of even $1$~s, one would need $\Delta\approx 3\times10^8$~m.

%------------------------------------------------------------------
\section{Summary and discussion}\label{sect:conclusion}
%------------------------------------------------------------------

We have described the causal structure of both eternal and dynamical warp drives. In a geometric optics approximation we have studied the propagation of light rays in dynamical geometries and found that the same exponential relation between affine coordinates on \scrim~and \scribub~is recovered at late times (large $u$) as in the case of black hole spacetimes. Given this relation it is unavoidable the conclusion that indeed a Hawking flux will be observed by any observer inside the warp-drive bubble far from the black horizon. Indeed, the calculation shows explicitly the onset of such a flux. This radiation is produced at the black horizon and soon fills the interior of the bubble, traveling rightward at the speed of light. The central region of the warp drive behaves like the asymptotic region of a black hole: In both these regions the static contribution ($\rho_{\rm st}$) to the energy density vanishes so that the total energy density is due solely to the Hawking radiation generated at the black horizon.

When creating a warp drive one not only forms a black horizon but also a white one. Both are sources of right-going radiation. To understand better the nature of this radiation we have calculated the RSET in this geometry and, in particular, the energy density as measured by freely falling observers. In this way we recover that the RSET does not diverge at the horizons at any finite time. The singular behavior of the static terms (or vacuum polarization terms) in the RSET at the horizons is canceled by the leading contributions of the dynamical right-going terms, or what is equivalent, by the presence of Hawking radiation at both the horizons. 

{It is however easy to see that the subleading terms of Eq.~\eqref{eq:sublead} behavior is rather different between the black and the white horizons. The subleading term in the RSET associated with the formation of the black horizon does not produce any significative backreaction on the horizon itself. In fact, this term is just a transient which decays exponentially.\footnote{Furthermore, the analogy with a black hole originated by a star collapse of \cite{stresstensor} also allows us to infer that the warp drive must be created very rapidly in order to avoid a huge accumulation of vacuum polarization at the horizon and so a huge initial value of the energy density.}

The formation of a white horizon is also associated with a similar subleading term,} but this time it accumulates onto the white horizon itself. This causes the energy density $\rho$ seen by a free-falling observer to grow unboundedly with time on this horizon. The semiclassical backreaction of the RSET will make the superluminal warp drive become rapidly unstable, in a time scale of the order of $1/\kappa_2$, the inverse of the surface gravity of the white horizon. Indeed, if one trusted the QI \cite{pfenningford,broeck}, the wall thickness for $v_0\approx c$ would be $\Delta\lesssim 10^2\,L_P$, and the surface gravity $\kappa\gtrsim10^{-2}\,{t_P}^{-1}$, where $t_P$ is the Planck time.\footnote{{In $1+1$ dimensions the warp-drive configuration is actually a vacuum solution of Einstein equations. In this case, QI will not impose any conditions on the size of the wall thickness. However, as we will show, we expect our results to be valid also in $3+1$ dimensions. Therefore, we use the wall thickness of $3+1$ warp drives to obtain a realistic estimation for the Hawking temperature and the time scale of the exponential growing of the RSET.}}
This means that the time scale over which the backreaction of the RSET would become important is $\tau\sim1/\kappa\lesssim10^2\, t_P$. Indeed, even forgetting about the QI, in order to get even a time scale $\tau\sim1 $ s for the growing rate of the RSET, one would need a wall as large as $3\times10^8$ m. Therefore, most probably one would be able to maintain a superluminal speed for just a very short interval of time. 
{In addition to the above mentioned growing {term} on $\hor_2^{-}$ we have shown that there is also an unbounded accumulation of Hawking radiation on \scribub. Also this contribution will very rapidly lead to a significant backreaction on the superluminal warp drive and to some sort of semiclassical instability of the solution (that will most probably prevent the formation of the Cauchy horizons at late times).

Interestingly, recent investigations~\cite{parentani} seem to imply that the above found asymptotic divergences of the RSET on the white horizon might disappear if Lorentz symmetry gets broken at high energies. While in this case we still expect that the RSET will acquire large values soon after the white horizon is formed, we do not know if this would be enough to prevent the sustainability of the warp drive. We think that this issue could be subject for future research perhaps within the context of analog models of gravity (see~\cite{lrr-2005-12} for a complete review on analog models) where it might even be addressed experimentally.

Even if the above described semiclassical instability could be avoided by some external action on the warp-drive bubble (or by some appropriate UV completion of the quantum field theory, like in~\cite{parentani}), the QI lead to the conclusion  that the Hawking radiation in the center of the bubble will burn the internal observer with an excruciating temperature of $T_H\sim\kappa\gtrsim10^{-2}\,T_P$, where $T_P$ is the Planck temperature, about $10^{32}$ K. This would prevent the use of a superluminal warp drive for any kind of practical purpose. If we do not trust the QI, this high temperature might be avoided by making thicker walls. For instance, with $\Delta\sim 1$ m, one obtains a temperature of about $0.003$ K (roughly the temperature of radiation at a wavelength of $1$ m).\footnote{However, some very effective taming for the growing backreaction at the white horizon would be needed also in this case, given the previously estimated RSET growing rate.
}
% \footnote{Note however, that still some very effective taming for the growing backreaction at the white horizon would be needed also in this case. In fact, in order to get even a time scale $\tau \sim 1$ s for the growing rate of the RSET at the white horizon, one would need a wall thickness as large as $3\times10^8$ m.}
}
%However, in our case, there is also a white horizon acting as a source of right-going radiation. However this radiation, being right-going, neither can travel from \hm~to the center of the bubble, nor can cross \hm, but it is accumulated onto the white horizon itself. This causes the energy density $\rho$ to grow unboundedly with time on this horizon.  {Moreover, we found that the Hawking radiation produced at the black horizon, causes $\rho$ to diverge close to the white horizon as time increases, because of a blue-shift effect}. Thus, as time grows, the RSET will be no more negligible and will give rise to a strong backreaction. We conclude that such a spacetime is likely to be very unstable because of this phenomenon. 

Finally we want to comment on the fact that in this paper a 1+1 calculation was performed. Generally in spherically symmetric spacetimes this could be seen as an $s$-wave approximation to the correct results. However, this is not the case for the axisymmetric warp-drive geometry. However, we do expect that the salient features of our results would be maintained in a full 3+1 calculation (most probably a numerical one) given that they will still be valid in a suitable open set of the horizons centered around the axis aligned with the direction of motion.  

In conclusion, we think that this work is casting strong doubts about the semiclassical stability of superluminal warp drives. Of course, all the aforementioned problems disappear when the bubble remains subluminal. In that case no horizons form, no Hawking radiation is created, and neither strong temperature nor white horizon instability is found. The only remaining problem is that one would still need the presence of some amount of exotic matter to maintain the subluminal drive. 
%Apart from this issue traveling at just 99\% of the speed of light could not be that bad after all.

\begin{acknowledgments}
The authors wish to thank S.~Sonego and M.~Visser for illuminating discussions. S.F. acknowledges the support provided by a INFN/MICINN collaboration. C.B. has been supported by the Spanish MICINN under Project No.~FIS2008-06078-C03-01/FIS and Junta de Andalucia under Project No.~FQM2288.
\end{acknowledgments}

\appendix

%-----------------------------------------------------------
\section{An example of dynamical warpdrive}\label{app:kink}
%-----------------------------------------------------------

In this Appendix we want to present a model of dynamical warpdrive, which we used for numerical calculation. It is particularly useful because it allows one to carry out an almost complete analytical treatment. We adapt here the method presented in~\cite{particlecreation,notrap} for stellar collapses to the creation of a warp drive from Minkowski spacetime.
We can choose a simplified piecewise velocity profile, which has the relevant properties of a dynamical warp drive, i.e., {it describes} a flat geometry at early times and coincides with {the metric in} Eq.~\eqref{eq:fluidmetric} after some finite time~$t>0$.
%
%Fortunately, the fundamental points are that a future apparent horizon is formed at a finite time $t_H$ and that at early times the spacetime is flat. These are the only two requirements such that $u$, (label of outgoing null rays at \scribub) is divergent when $x$ approaches \hp, and $U$ (label of standard plane waves at \scrim) is regular on \hp. Thus, we try to find a metric in the form of Eq. (\ref{eq:ppvelocity}) which  is flat
%Of course, this choice is made only to have a computable model, however let us stress that the the so obtained RSET will be relevant also  
%Then we perform our calculation just like in Ref.~\onlinecite{particlecreation} and our result is correct also 
%for the metric of Eq. (\ref{eq:dynamicfluidmetric}) given that 
%In fact we obtain a result of the form $U=A+B e^{-\kappa u}$ with $\kappa $ the surface gravity. The result with the metric of Eq. (\ref{eq:dynamicfluidmetric}) could differ in the two constants $A$ and $B$, but the leading term must go in the same way $U\propto e^{-\kappa u}$ when approaching \hp, with the same surface gravity and the same Hawking temperature. In an analogous way we can study what happens to rays passing close to the white horizon.}
%We have to find a suitable metric, flat for $t\to-\infty$, with the kinklike structure of Eq. (\ref{eq:ppvelocity}), such that the warp-drive spacetime is uncovered at finite time $t_H$. 

Using our velocity profile defined in Eq.~\eqref{eq:fluidvelocity}
\begin{equation}
 \bar{v}(r)=\alpha c\left[f(r)-1\right]~,
\end{equation}
we can define a dynamical profile by replacing $\hat{v}$ in Eq.~\eqref{eq:dynamicfluidmetric} by $\hat{v}_{\rm kink}$
\begin{equation}\label{eq:ppwarpvelocity}
 \hat{v}_{\rm kink}(r,t)=
 \left\{
  \begin{aligned}
    & \bar{v}(\xi(t)) \qquad&\text{if} \quad |r|\geq\xi(t)~,  \\
    & \bar{v}(r) \qquad&\text{if}\quad |r|<\xi(t)~,
  \end{aligned}
 \right.
\end{equation}
where $\xi(t)$ is a monotonically increasing function of $t$, such that $\xi(t)\to0$, for $t\to-\infty$, and $\xi(t_H)=r_2=-r_1$.

One may wonder whether defining a velocity profile with a kink, as in Eq. \eqref{eq:ppwarpvelocity},  may lead to unphysical phenomena. Indeed this computational trick induces some spurious effects, but
%we shall see below that
these features are just transients and do not affect the results at late times.

%----------------------------------------------
\subsection{Right-going rays}
%----------------------------------------------

We apply the same procedure of \cite{particlecreation} to calculate the exact relation between the past null coordinate
$U$ in $\scri_L^-$ and the future null coordinate relevant at \scribub.
%In the dynamical warp-drive there is a single past null coordinate but three different future null coordinates associated with the final regions $I$, $II$ and $III$, as described before. Here on we will be dealing exclusively with the connection between $\scri_L^-$ and the interior of the bubble. Therefore we will use $u$ to denote $u_{\rm II}$ whenever this does not lead to confusion. As discussed in \cite{particlecreation} the relation $U=p(u)$ encodes all the relevant information about particle production.
To find this relation one has to find the integral curves of the ray differential equation~\eqref{eq:rays}, which becomes at early times
\begin{equation}
 \frac{dr}{dt}=c
\end{equation}
and in a neighborhood of \scribub~($t\to +\infty$ and $r\to r_2$, see Fig.~\ref{fig:dynwd-pen})	
\begin{multline}\label{eq:rayslimit}
 \frac{dr}{dt}=c+\bar{v}(r)
 %\approx c+\left[-c+{\left.\frac{dv}{dr}\right|}_{r=r_2}\left(r-r_2\right)\right]\\
	=\kappa \left(r_2-r\right)+{\cal O}\left((r_2 -r)^2\right).
\end{multline}
Integrating the first equation we obtain the obvious result
\begin{equation}
 t=C+\frac{r}{c}
\end{equation}
and for the second one
\begin{equation}\label{eq:laterays}
 t=D-\frac{1}{\kappa }\ln\left(r_2-r\right)~.
\end{equation}
Following \cite{particlecreation} we identify initial events $P\equiv\left(r_i,t_i\right)$, with $r_i\sim ct_i$, final events $Q\equiv\left(r_f,t_f\right)$ with $r_f\sim r_2-e^{-\kappa t_f}$, and intermediate events $O\equiv\left(r_0,t_0\right)$.

Let us define
\begin{align}
    U &= \lim_{t_i\to-\infty}\left(t_i-\frac{r_i}{c}\right)~,\label{eq:Ulimit}\\
    u &= \lim_{t_f\to+\infty}\left[t_f+\frac{1}{\kappa }\ln\left(r_2-r_f\right)\right]~.\label{eq:ulimit}
\end{align}
Integrating Eq.~\eqref{eq:rays} between $P$ and an intermediate event $O$ in the in region of this spacetime (where the velocity profile depends only on $t$), we find
\begin{equation}\label{eq:PO1}
 r_0-r_i=\int_{t_i}^{t_0}dt\left[c+\bar{v}(\xi(t))\right]~.
\end{equation}
Using the definition of $U$ in Eq.~\eqref{eq:Ulimit} we obtain
\begin{equation}\label{eq:Utr}
 U=t_0-\frac{r_0}{c}+\frac{1}{c}\int_{-\infty}^{t_0}dt\bar{v}(\xi(t))~,
\end{equation}

Then, integrating between an intermediate event $O$ now in the out region (where the velocity profile depends only on $r$) and $Q$ we find
\begin{equation}\label{eq:O2Q}
 t_f-t_0=\int_{r_0}^{r_f}\frac{dr}{c+\bar{v}(r)}~,
\end{equation}
\begin{widetext}
and using the definition in Eq.~\eqref{eq:ulimit}
\begin{multline}
 u=t_0+\lim_{r_f\to r_2}\left[\frac{1}{\kappa }\ln\left(r_2-r_f\right) + \int_{r_0}^{r_f}\frac{dr}{c+\bar{v}(r)}\right]\\
%=t_0+\lim_{r_f\to r_2}\left\{\frac{1}{\kappa }\ln\left(r_2-r_f\right)+ \int_{r_0}^{r_f}\frac{dr}{\kappa \left(r_2-r\right)}+\int_{r_0}^{r_f}dr\left[\frac{1}{c+\bar{v}(r)}-\frac{1}{\kappa \left(r_2-r\right)}\right] \right\}\\
=t_0+\frac{1}{\kappa}\ln\left[r_2-r_0\right]
-\frac{1}{\kappa}\ln\left[r_0-r_1\right]
+\frac{1}{\kappa}\ln\left[r_2-r_1\right]
+\lim_{r_f\to r_2} \int_{r_0}^{r_f}dr
\left[\frac{1}{c+\bar{v}(r)}-\frac{1}{\kappa \left(r_2-r\right)}-\frac{1}{\kappa \left(r-r_1\right)}\right]~.
\end{multline}
It is easy to see that the limit in the previous expression is finite for whatever $r_1<r_0<r_2$ so that we end up with the relation
\begin{equation}\label{eq:utr}
u=t_0+\frac{1}{\kappa}\ln\left[r_2-r_0\right]
-\frac{1}{\kappa}\ln\left[r_0-r_1\right]
+\frac{1}{\kappa}\ln\left[r_2-r_1\right]
+\int_{r_0}^{r_2}dr
\left[\frac{1}{c+\bar{v}(r)}-\frac{1}{\kappa \left(r_2-r\right)}-\frac{1}{\kappa \left(r-r_1\right)}\right]~.
\end{equation}
\end{widetext}

We now want to find the relation between $U$ and $u$. It is possible to find a particular form for the kink function $\xi$ (see Fig.~\ref{fig:lightrayssimp} and its caption for an example), such that all the rays cross either the 
left kink $r=-\xi(t)$ or the right kink $r=+\xi(t)$ only once. The event "crossing the kink" can be seen as belonging to both the in region and the out region. Therefore the relation $U=p(u)$ can be found by eliminating $t_0$ and $r_0$ between expressions \eqref{eq:Utr} and \eqref{eq:utr} taking into account that $r_0$ will be either $-\xi(t_0)$ or $+\xi(t_0)$. Figure~\ref{fig:lightrayssimp} shows an exact numerical calculation of ray propagation, performed with this method.

However, in order to study the late-time behavior of both particle production and the RSET, we only need the previous relation (see Sec.~\ref{sect:radiation}) for large and positive values of $u$ (the location of the black horizon) and for large and negative values of $u$ (the location of the white horizon). We want to show here that our particular example leads in fact to the general relation Eq.~\eqref{eq:Uu}. In this specific case, taking the limit for $|u|\to\infty$ corresponds to study the previous relation for light rays crossing the left kink at $r_0=-\xi(t_0)$, when $r_0$ is very close to $r_1$ [respectively, crossing the right kink at $r_0=+\xi(t_0)$, when $r_0$ is very close to $r_2$].
%As a consequence we can make $O_1$ and $O_2$ coinciding at a certain point $O=(t_0,r_0)$ where the ray crosses the kink. Then, we have one more relation $r_0=-\xi(t_0)$ (respectively, $r_0=+\xi(t_0)$). Eqs. (\ref{eq:Utr}) and (\ref{eq:utr}) become in this case:
%\begin{equation}
% U_\pm =t_0\pm\frac{\xi(t_0)}{c}+\frac{1}{c}\int_{-\infty}^{t_0}dt\,\bar{v}(\xi(t)),\label{eq:Uxi}\\
%\end{equation}
%\begin{multline}
% u_\pm =t_0+\frac{1}{\kappa }\ln\left[r_2\pm\xi(t_0)\right]\\
%	+\int_{\mp\xi(t_0)}^{r_2}dr\frac{\kappa \left(r_2-r\right)-\left[c+\bar{v}(r)\right]}{\kappa \left[c+\bar{v}(r)\right]\left(r_2-r\right)},\label{eq:uxi}
%\end{multline}
%where the upper (respectively, lower) sign is for rays passing close to the black (respectively, white) horizon.
%We can determine the relation between $U$ and $u$ close to the horizons, in the limit $\mp\xi(t_0)\to r_\pm$. 
First note that, when $t\to t_H$ (the time of the first appearance of the trapping horizons) the function $\xi$ can be expanded in the following form:
\begin{align}\label{eq:xi}
 \xi(t)&=r_2+\lambda\left(t-t_H\right)+{\cal O}\left[{(t-t_H)}^2\right]~,\\
 -\xi(t)&=r_1-\lambda\left(t-t_H\right)+{\cal O}\left[{(t-t_H)}^2\right]~,
\end{align}
with $\lambda$ a positive constant (remember that we have chosen for simplicity $r_1=-r_2$).
%Let us study Eq. (\ref{eq:Uxi}) for $t_0=t_H$:
Defining
\begin{equation}
%U_{H\pm}
U_{\substack{\rm BH\\\rm WH}}
=t_H\pm\frac{\xi(t_H)}{c}+\frac{1}{c}\int_{-\infty}^{t_H}dt\bar{v}(\xi(t))~,
\end{equation}
it is easy to see that
\begin{equation}
 U_\pm=U_{\substack{\rm BH\\\rm WH}}\mp \frac{\lambda}{c}\left(t_H-t_0\right)+{\cal O}\left[{(t_H-t_0)}^2\right]~,
\end{equation}
where with $U_+$ we denote the form of the function $U(t_0)$ for rays crossing the kink close to $r_1$ [respectively, with $U_-$ we denote the form of the function $U(t_0)$ for rays crossing the kink close to to $r_1$].
%\begin{widetext}
%
%and for $t_0\to t_H^-$:
%\begin{multline}
% U_\pm=U_{H\pm}+t_0-t_H\pm\frac{\xi(t_0)-\xi(t_H)}{c}-\frac{1}{c}\int_{t_0}^{t_H}dt\,\bar{v}(\xi(t))\\
%  =U_{H\pm}+t_0-t_H\pm\frac{\xi(t_0)\pm r_\pm}{c}
%	-\frac{1}{c}\int_{t_0}^{t_H}dt\left\{-c+{\left.\frac{dv}{dr}\right|}_{r=r_\pm}\left(\mp\xi(t)-r_\pm\right)+O\left[{(\mp\xi(t)-r_\pm)}^2\right]\right\}\\
%  =U_{H\pm}\pm\frac{\lambda\left(t_0-t_H\right)+O\left[{(t_0-t_H)}^2\right]}{c}+\frac{1}{c}\int_{t_0}^{t_H}dt \left\{\kappa \lambda\left(t-t_H\right)+O\left[{(t-t_H)}^2\right]\right\}\\
%  =U_{H\pm}\mp\frac{\lambda}{c}\left(t_H-t_0\right)+O\left[{(t_H-t_0)}^2\right].
%\end{multline}
%\end{widetext}
%
By expanding Eq.~\eqref{eq:utr} in the same limit $t_0\to t_H$ and retaining only the dominant term we obtain
\begin{equation}
 u_\pm\simeq\mp\frac{1}{\kappa }\ln\left[\lambda\left(t_H-t_0\right)\right]~,
\end{equation}
where we define $u_\pm$ in the same fashion as $U_\pm$.
%
%\begin{widetext}
%\begin{align}
%\begin{split}
% u_+&\sim \int_{-\xi(t_0)}^{r_2}dr\frac{\kappa \left(r_2-r\right)-\left[c+\bar{v}(r)\right]}{\kappa \left[c+\bar{v}(r)\right]\left(r_2-r\right)}\sim \int_{-\xi(t_0)}^{r_2}dr\frac{\kappa \left(r_2-r_1\right)-\left[c+(-c)+{\left.\frac{dv}{dr}\right|}_{r=r_1}\left(r-r_1\right)\right]}{\kappa \left[c+(-c)+{\left.\frac{dv}{dr}\right|}_{r=r_1}\left(r-r_1\right)\right]\left(r_2-r_1\right)}\\
%  &\sim \int_{-\xi(t_0)}^{r_2}dr\frac{1}{\kappa \left(r-r_1\right)}=\frac{1}{\kappa }\ln\left[\frac{r_2-r_1}{-\xi(t_0)-r_1}\right]\sim -\frac{1}{\kappa }\ln\left[\lambda\left(t_H-t_0\right)\right],
%\end{split}\\
%\begin{split}
% u_-&\sim+\frac{1}{\kappa }\ln\left[r_2-\xi(t_0)\right]=\frac{1}{\kappa }\ln\left[\lambda\left(t_H-t_0\right)\right].
%\end{split}
%\end{align}
%\end{widetext}
%
Putting together the last two results we obtain (inside the bubble) the general relation Eq.~\eqref{eq:Uu}:
\begin{equation}
 U(u \to \pm \infty)=U_{\substack{\rm BH \\ \rm WH}} \mp A_{\pm} e^{\mp\kappa u}.
\end{equation}

\subsection{Left-going rays}

In Sec.~\ref{sect:RSET} we stated that it is possible to find a wide class of transitions from a Minkowskian geometry to a warp-drive one such that the contributions to the RSET due to left-going modes are just transient, vanishing at late times. We show here that this is the case for our specific model.

In order to do so, we shall need the relation between ingoing and outgoing left-going rays. Left-going rays are the solution of the differential equation~\eqref{eq:raysleft} which becomes at early times
\begin{equation}
 \frac{dr}{dt}=-c~,
\end{equation}
while inside the bubble we cannot take any limit as in Eq.~\eqref{eq:rayslimit} because left-going rays are not confined inside the bubble, but escape through the black horizon, as we already noticed in Sec.~\ref{sect:radiation}. Integrating the equation at early times we obtain
\begin{equation}
 t=C-\frac{r}{c}~.
\end{equation}
We identify initial events $P\equiv\left(r_i,t_i\right)$, with $r_i\sim ct_i$, final events $Q\equiv\left(r_f,t_f\right)$, and intermediate events $O\equiv\left(r_0,t_0\right)$ either in the in region (when the velocity profile depends only on $t$) or in the out region (when the velocity profile depends only on $r$).

Using $\tilde w$ defined as in Eq.~\eqref{eq:wdef}, choosing arbitrarily the constant of integration, we obtain
%However we do not have a natural definition of the coordinate $w$ for left-going rays as in Eq. (\ref{eq:ulimit}).  However, what we need is a null coordinate such that, for each ray, $w$ is an integration constant of Eq. (\ref{eq:raysleft}) beyond the kink, where $v$ depends only on $r$. From this point of view we see that Eqs. (\ref{eq:Ulimit}), (\ref{eq:ulimit}) and (\ref{eq:Wlimit}) determine the zero-point respectively of $U$, $u$, $W$ in a natural way. For $W$ this is not possible, so we decide arbitrarily its zero-point. We define
\begin{equation}\label{eq:w}
 \tilde w(t_0,r_0)\equiv t_0+\int_0^{r_0}\frac{dr}{c-\bar{v}(r)}~.
\end{equation}

For $W$ we proceed as for $U$. We integrate Eq.~\eqref{eq:raysleft} between $P$ and an event $O$ in the in region
\begin{equation}\label{eq:PO1left}
 r_0-r_i=\int_{t_i}^{t_0}dt\left[-c+\bar{v}(\xi(t))\right]~.
\end{equation}
Defining $W$ as
\begin{align}
    W &= \lim_{t_i\to-\infty}\left(t_i+\frac{r_i}{c}\right)~,\label{eq:Wlimit}
\end{align}
we obtain
\begin{equation}\label{eq:Wtr}
 W=t_0+\frac{r_0}{c}-\frac{1}{c}\int_{-\infty}^{t_0}dt\bar{v}(\xi(t))~.
\end{equation}
Now, the relation $W=q(\tilde w)$ can be found eliminating $t_0,r_0$ from Eqs.~\eqref{eq:wdef} and \eqref{eq:Wtr} in the limit where $(t_0,r_0)$ is on the kink.

We can now show that indeed the term $\rho_{{\rm dyn}-\tilde{w}}$ defined in Eq.~\eqref{eq:rhodw} is just a transient for this model. We see from Fig.~\ref{fig:lightrayssimp} that, as time grows, points inside the bubble are reached by $\tilde{w}$ rays that intersect the kink later and later. Using Eqs.~\eqref{eq:wdef} and \eqref{eq:Wtr} we can estimate $\dot{q}$, calculating the derivatives of $\tilde{w}$ and $W$ with respect to $t$ on the kink, at crossing time $t_c$:
\begin{multline}
 \dot{q}=\frac{dW}{d\tilde{w}}=\left.\frac{dW}{dt}\right|_{r=\xi(t_c)}\left(\left.\frac{d\tilde{w}}{dt}\right|_{r=\xi(t_c)}\right)^{-1}\\
	=\left(1+\frac{\dot\xi(t_c)}{c}-\frac{\bar{v}(\xi(t_c))}{c}\right)\left(1+\frac{\dot\xi(t_c)}{c-\bar{v}(\xi(t_c))}\right)^{-1}\\
 =1-\frac{\bar{v}(\xi(t_c))}{c}\to 1-\frac{\bar{v}(+\infty)}{c}~.
\end{multline}
This is a constant value greater than $0$.
% Moreover $\dot{q}$ tends exponentially fast to this value, given a shift-velocity profile as in Eq. (\ref{eq:fluidvelocity}).
As a consequence, all the derivatives of $\dot{q}$ go to zero as time grows and $\rho_{{\rm dyn}-\tilde{w}}$ must go to zero in the same way. In conclusion, the dynamic term originated by the distortion of left-going rays can be different from zero when the horizon is created. However, this is only a transient term that is brought toward $\scri_L^+$, i.e., outside the bubble. That is, it is possible to create a transition region such that some radiation traveling leftward is produced only at the {onset} of the warp drive, but there is not any phenomenon like Hawking radiation originated in this way.

%-------------------------------------------------------
\section{Null coordinates}\label{app:nullcoordinates}
%-------------------------------------------------------

In this Appendix we summarize some results concerning systems of null coordinates. In Sec.~\ref{sect:radiation} we defined two sets of null coordinates $(U,W)$ and $(u,\tilde{w})$ obtained by integrating equations of light propagation, respectively, in the in region, in which the metric is Minkowskian, and in the out region, in which it depends on $r$ (after the kink, in the particular example of Appendix~\ref{app:kink}). Now, let us rewrite the metric using the set of coordinates $(u,\tilde{w})$.
%We calculate the derivatives of this coordinates with respect to $r$ and $t$.
%
\comment{
Before the kink ($|r|>\xi(t)$), using Eqs.~\eqref{eq:Utr} and \eqref{eq:Wtr}:
\begin{equation}\label{eq:UWderivs}
begin{aligned}
  U_r &= -\frac{1}{c},&\quad W_r &=\frac{1}{c},\\
  U_t &= 1+\frac{\bar{v}\left(\xi(t)\right)}{c},&\quad W_t &=1-\frac{\bar{v}\left(\xi(t)\right)}{c},
\end{aligned}
\end{equation}
}
In the out region, using Eqs.~\eqref{eq:udef} and \eqref{eq:wdef}:
\begin{equation}\label{eq:uwderivs}
\begin{aligned}
  u_r &= -\frac{1}{c+\bar{v}(r)}~,	&\quad		\tilde{w}_r &=\frac{1}{c-\bar{v}(r)}~,\\
  u_t &= 1~,				&\quad		\tilde{w}_t &=1~.
\end{aligned}
\end{equation}
\comment{
Now we put these expressions into Eqs.~\eqref{eq:metricUW} and \eqref{eq:metricuw}.
\begin{multline}
  ds^2=-C(U,W)dUdW \\
	=-C(U,W)\left(U_t dt + U_r dr\right)\left(W_t dt + W_r dr\right)\\
	=\frac{C(U,W)}{c^2}\left\{c^2 dt^2-\left[dr-\bar{v}\left(\xi(t)\right) dt\right]^2\right\}~,
\end{multline}
if $|r|>\xi(t)$, and
}
Now we put these expressions into Eq.~\eqref{eq:metricuw}:
\begin{multline}
  ds^2=-\bar{C}(u,\tilde{w})dud\tilde{w} \\
	=-\bar{C}(u,\tilde{w})\left(u_t dt + u_r dr\right)\left(\tilde{w}_t dt + \tilde{w}_r dr\right)\\
   	=\frac{\bar{C}(u,\tilde{w})}{c^2-\bar{v}(r)^2}\left\{c^2 dt^2-\left[dr-\bar{v}(r) dt\right]^2\right\}~.
\end{multline}
%if $|r|<\xi(t)$.\\
By comparison with Eq.~\eqref{eq:dynamicfluidmetric} we obtain
\comment{
\begin{align}
 C(U,W)=1,\qquad&\text{if }|r|>\xi(t)~,\label{eq:C}\\
 \bar{C}(u,\tilde{w})=c^2-\bar{v}(r)^2,\qquad&\text{if }|r|<\xi(t)~,\label{eq:barC}
\end{align}
}
\begin{equation}
 \bar{C}(u,\tilde{w})=c^2-\bar{v}(r)^2~,\label{eq:barC}
\end{equation}

Note that the definition of $U$ and $W$ {can be extended} in the out region by following the light rays coming from the in region (in the model of Appendix~\ref{app:kink} by matching on the kink the solutions of $U$ and $W$ with those of $u$ and $\tilde{w}$). For instance, one can {naturally} define $U(t,r)$ in the out region {as} $U(t,r)=p\left(u(t,r)\right)$ and in analogous ways for the other null coordinate. This means that it makes sense to take derivatives of the null coordinates with respect to $r$ and $t$ in all the spacetime. We write here the derivatives of $U$ and $W$ in the out region:
\begin{equation}\label{eq:UWderivspq}
\begin{aligned}
  U_r &=\dot{p}(u)u_r = -\frac{\dot{p}(u)}{c+\bar{v}(r)}~,	&\quad		W_r &=\dot{q}(\tilde{w})\tilde{w}_r =\frac{\dot{q}(\tilde{w})}{c-\bar{v}(r)}~,\\
  U_t &=\dot{p}(u)u_t = \dot{p}(u)~,				&\quad		 W_t &=\dot{q}(\tilde{w})\tilde{w}_t =\dot{q}(\tilde{w})~.
\end{aligned}
\end{equation}

%--------------------------------------------------------------
\section{Asymptotic expansion of $f(u)$}\label{app:fp}
%--------------------------------------------------------------
%
\comment{
We know from Eq.~\eqref{eq:uwapprox} that $u$ diverges logarithmically with $r$ close to the horizon:
% or, in equivalent ways:
\begin{gather}
 u\simeq t\mp\frac{1}{\kappa}\ln\left|r-r_{1,2}
 %{\substack{1 \\ 2}}
\right|,\label{eq:uwhapprox}
 %\\
 %\left|r-r_{1,2}
 %{\substack{1 \\ 2}}
\right|\propto e^{\mp ku}.\label{eq:xprop}
\end{gather}
And our coordinate $U$ is regular on the horizons, so we can write it as
\begin{equation}
 U_\pm=\mathcal{U}_\pm\left(r-r_{1,2}
 %{\substack{1 \\ 2}}
\right),
\end{equation}
where $\mathcal{U}\pm$ are analytic functions. Inserting Eq.~\eqref{eq:uwhapprox} in this expression
\begin{equation}
 U_\pm=p(u_\pm)=\mathcal{P}_\pm(e^{\mp\kappa u_\pm}),
\end{equation}
}
We know from Sec.~\ref{sect:radiation} that $U$ can be expanded for large absolute values of $u$ in a Taylor series in $e^{\mp\kappa u}$. In fact we found that
\begin{equation}
 U_\pm=p(u\to\pm \infty)=\mathcal{P}_\pm(e^{\mp\kappa u})~,
\end{equation}
where $\mathcal{P}_\pm$ are analytic functions. We now expand it up to the third order:
\begin{multline}\label{eq:Uulong}
 U=p(u \to \pm \infty)= U_{\substack{\rm BH \\ \rm WH}}+A_{1\pm} e^{\mp\kappa u}\\
	+\frac{A_{2\pm}}{2} e^{\mp 2\kappa u}+\frac{A_{3\pm}}{6}e^{\mp 3\kappa u}+{\cal O}\left(e^{\mp 4\kappa u}\right)~,
\end{multline}
where the upper (lower) signs correspond to $u\to+\infty$ ($u\to -\infty$). For simplicity we have defined the coefficients $A_{1+}\equiv-A_+$ and $A_{1-}\equiv A_-$.
In order to calculate the stress-energy tensor we need to calculate the term Eq.~\eqref{eq:fu}
\begin{equation}
f(u)=\frac{3\ddot{p}^2-2\dot{p}\,\dddot{p}}{\dot{p}^2}~.
\end{equation}

%Let us start by looking at the behavior of $f(u)$ at late times (large $u$) 
We have
\begin{widetext}
\begin{gather}
 \dot{p}(u)=\mp\kappa A_{1\pm} e^{\mp\kappa u}\left[1+\frac{A_{2\pm}}{A_{1\pm}} e^{\mp\kappa u}+\frac{A_{3\pm}}{2A_{1\pm}}e^{\mp 2\kappa u}+{\cal O}\left(e^{\mp 3\kappa u}\right)\right]~,\label{eq:approxpu}\\
 \ddot{p}(u)=\kappa^2 A_{1\pm} e^{\mp\kappa u}\left[1 +\frac{2A_{2\pm}}{A_{1\pm}} e^{\mp\kappa u}+\frac{3 A_{3\pm}}{2A_{1\pm}}e^{\mp 2\kappa u}+{\cal O}\left(e^{\mp 3\kappa u}\right)\right]~,\\
 \dddot{p}(u)=\mp\kappa^3 A_{1\pm} e^{\mp\kappa u}\left[1 +\frac{4A_{2\pm}}{A_{1\pm}} e^{\mp\kappa u}+\frac{9 A_{3\pm}}{2A_{1\pm}}e^{\mp 2\kappa u}+{\cal O}\left(e^{\mp 3\kappa u}\right)\right]~.
\end{gather}
so that
\begin{multline}\label{eq:approxfu}
f(u)=\frac{\kappa^4 A_{1\pm}^2e^{\mp 2\kappa u}
		\left\{1+2{\left(A_{2\pm}/A_{1\pm}\right)}e^{\mp\kappa u}+\left[4{\left(A_{2\pm}/A_{1\pm}\right)}^2-A_{3\pm}/A_{1\pm}\right]e^{\mp 2\kappa u}+{\cal O}\left(e^{\mp 3\kappa u}\right)\right\}}
	{\kappa^2 A_{1\pm}^2e^{\mp 2\kappa u}
		\left\{1+2{\left(A_{2\pm}/A_{1\pm}\right)}e^{\mp\kappa u}+\left[{\left(A_{2\pm}/A_{1\pm}\right)}^2+A_{3\pm}/A_{1\pm}\right]e^{\mp 2\kappa u}+{\cal O}\left(e^{\mp 3\kappa u}\right)\right\}}\\
=\kappa^2\left\{1+\frac{\left[4{\left(A_{2\pm}/A_{1\pm}\right)}^2-A_{3\pm}/A_{1\pm}\right]}{1+2{\left(A_{2\pm}/A_{1\pm}\right)}e^{\mp\kappa u}}e^{\mp 2\kappa u}+{\cal O}\left(e^{\mp 3\kappa u}\right) \right\}
	\times\left\{1-\frac{\left[{\left(A_{2\pm}/A_{1\pm}\right)}^2+A_{3\pm}/A_{1\pm}\right]}{1+2{\left(A_{2\pm}/A_{1\pm}\right)}e^{\mp\kappa u}}e^{\mp 2\kappa u}+{\cal O}\left(e^{\mp 3\kappa u}\right) \right\}\\
=\kappa^2\left\{1+\left[3{\left(\frac{A_{2\pm}}{A_{1\pm}}\right)}^2-2\frac{A_{3\pm}}{A_{1\pm}}\right]e^{\mp 2\kappa u}+{\cal O}\left(e^{\mp 3\kappa u}\right) \right\}~,
\end{multline}
and we can finally use the expansion {of} Eq.~\eqref{eq:uwhapprox}, valid for points close to the horizons, to obtain
\begin{equation}\label{eq:approxf}
 f(u) =\kappa^2\left\{1+\left[3{\left(\frac{A_{2\pm}}{A_{1\pm}}\right)}^2-2\frac{A_{3\pm}}{A_{1\pm}}\right]e^{\mp 2\kappa t}\left(r-r_{1,2}
 %{\substack{1 \\ 2}}
\right)^2+{\cal O}\left(\left(r-r_{1,2}
 %{\substack{1 \\ 2}}
\right)^3\right) \right\}~.
\end{equation}
\end{widetext}

\bibliographystyle{apsrev}
\bibliography{warpdrive}

\begin{thebibliography}{21}
\expandafter\ifx\csname natexlab\endcsname\relax\def\natexlab#1{#1}\fi
\expandafter\ifx\csname bibnamefont\endcsname\relax
  \def\bibnamefont#1{#1}\fi
\expandafter\ifx\csname bibfnamefont\endcsname\relax
  \def\bibfnamefont#1{#1}\fi
\expandafter\ifx\csname citenamefont\endcsname\relax
  \def\citenamefont#1{#1}\fi
\expandafter\ifx\csname url\endcsname\relax
  \def\url#1{\texttt{#1}}\fi
\expandafter\ifx\csname urlprefix\endcsname\relax\def\urlprefix{URL }\fi
\providecommand{\bibinfo}[2]{#2}
\providecommand{\eprint}[2][]{\url{#2}}

\bibitem[{\citenamefont{{Alcubierre}}(1994)}]{alcubierre}
\bibinfo{author}{\bibfnamefont{M.}~\bibnamefont{{Alcubierre}}},
  \bibinfo{journal}{Classical and Quantum Gravity}
  \textbf{\bibinfo{volume}{11}}, \bibinfo{pages}{L73} (\bibinfo{year}{1994}),
  \eprint{arXiv:gr-qc/0009013}.

\bibitem[{\citenamefont{{Lobo}}(2007)}]{lobo2007}
\bibinfo{author}{\bibfnamefont{F.~S.~N.} \bibnamefont{{Lobo}}},
  \bibinfo{journal}{ArXiv e-prints}  (\bibinfo{year}{2007}),
  \eprint{0710.4474}.

\bibitem[{\citenamefont{{Everett}}(1996)}]{everett}
\bibinfo{author}{\bibfnamefont{A.~E.} \bibnamefont{{Everett}}},
  \bibinfo{journal}{\prd} \textbf{\bibinfo{volume}{53}}, \bibinfo{pages}{7365}
  (\bibinfo{year}{1996}).

\bibitem[{\citenamefont{{Lobo} and
  {Visser}}(2004{\natexlab{a}})}]{lobovisser2004a}
\bibinfo{author}{\bibfnamefont{F.~S.~N.} \bibnamefont{{Lobo}}}
  \bibnamefont{and} \bibinfo{author}{\bibfnamefont{M.}~\bibnamefont{{Visser}}},
  \bibinfo{journal}{Classical and Quantum Gravity}
  \textbf{\bibinfo{volume}{21}}, \bibinfo{pages}{5871}
  (\bibinfo{year}{2004}{\natexlab{a}}), \eprint{arXiv:gr-qc/0406083}.

\bibitem[{\citenamefont{{Lobo} and
  {Visser}}(2004{\natexlab{b}})}]{lobovisser2004b}
\bibinfo{author}{\bibfnamefont{F.~S.~N.} \bibnamefont{{Lobo}}}
  \bibnamefont{and} \bibinfo{author}{\bibfnamefont{M.}~\bibnamefont{{Visser}}},
  \bibinfo{journal}{ArXiv General Relativity and Quantum Cosmology e-prints}
  (\bibinfo{year}{2004}{\natexlab{b}}), \eprint{gr-qc/0412065}.

\bibitem[{\citenamefont{{Pfenning} and {Ford}}(1997)}]{pfenningford}
\bibinfo{author}{\bibfnamefont{M.~J.} \bibnamefont{{Pfenning}}}
  \bibnamefont{and} \bibinfo{author}{\bibfnamefont{L.~H.}
  \bibnamefont{{Ford}}}, \bibinfo{journal}{Classical and Quantum Gravity}
  \textbf{\bibinfo{volume}{14}}, \bibinfo{pages}{1743} (\bibinfo{year}{1997}),
  \eprint{arXiv:gr-qc/9702026}.

\bibitem[{\citenamefont{{Roman}}(2005)}]{roman}
\bibinfo{author}{\bibfnamefont{T.~A.} \bibnamefont{{Roman}}}, in
  \emph{\bibinfo{booktitle}{The Tenth Marcel Grossmann Meeting. On recent
  developments in theoretical and experimental general relativity, gravitation
  and relativistic field theories}}, edited by
  \bibinfo{editor}{\bibfnamefont{M.}~\bibnamefont{{Novello}}},
  \bibinfo{editor}{\bibfnamefont{S.}~\bibnamefont{{Perez Bergliaffa}}},
  \bibnamefont{and} \bibinfo{editor}{\bibfnamefont{R.}~\bibnamefont{{Ruffini}}}
  (\bibinfo{publisher}{World Scientific Publishing},
  \bibinfo{address}{Singapore}, \bibinfo{year}{2005}), p.
  \bibinfo{pages}{1909}.

\bibitem[{\citenamefont{{Van Den Broeck}}(1999)}]{broeck}
\bibinfo{author}{\bibfnamefont{C.}~\bibnamefont{{Van Den Broeck}}},
  \bibinfo{journal}{Classical and Quantum Gravity}
  \textbf{\bibinfo{volume}{16}}, \bibinfo{pages}{3973} (\bibinfo{year}{1999}),
  \eprint{arXiv:gr-qc/9905084}.

\bibitem[{\citenamefont{{Hiscock}}(1997)}]{hiscock}
\bibinfo{author}{\bibfnamefont{W.~A.} \bibnamefont{{Hiscock}}},
  \bibinfo{journal}{Classical and Quantum Gravity}
  \textbf{\bibinfo{volume}{14}}, \bibinfo{pages}{L183} (\bibinfo{year}{1997}),
  \eprint{arXiv:gr-qc/9707024}.

\bibitem[{\citenamefont{{Barcel{\'o}} et~al.}(2008)\citenamefont{{Barcel{\'o}},
  {Liberati}, {Sonego}, and {Visser}}}]{stresstensor}
\bibinfo{author}{\bibfnamefont{C.}~\bibnamefont{{Barcel{\'o}}}},
  \bibinfo{author}{\bibfnamefont{S.}~\bibnamefont{{Liberati}}},
  \bibinfo{author}{\bibfnamefont{S.}~\bibnamefont{{Sonego}}}, \bibnamefont{and}
  \bibinfo{author}{\bibfnamefont{M.}~\bibnamefont{{Visser}}},
  \bibinfo{journal}{\prd} \textbf{\bibinfo{volume}{77}},
  \bibinfo{pages}{044032} (\bibinfo{year}{2008}).

\bibitem[{\citenamefont{{Gonz{\'a}lez-D{\'{\i}}az}}(2007)}]{gonzales2007}
\bibinfo{author}{\bibfnamefont{P.~F.}
  \bibnamefont{{Gonz{\'a}lez-D{\'{\i}}az}}}, \bibinfo{journal}{Physics Letters
  B} \textbf{\bibinfo{volume}{653}}, \bibinfo{pages}{129}
  (\bibinfo{year}{2007}).

\bibitem[{\citenamefont{{Barcel{\'o}} et~al.}(2004)\citenamefont{{Barcel{\'o}},
  {Liberati}, {Sonego}, and {Visser}}}]{causalstructure}
\bibinfo{author}{\bibfnamefont{C.}~\bibnamefont{{Barcel{\'o}}}},
  \bibinfo{author}{\bibfnamefont{S.}~\bibnamefont{{Liberati}}},
  \bibinfo{author}{\bibfnamefont{S.}~\bibnamefont{{Sonego}}}, \bibnamefont{and}
  \bibinfo{author}{\bibfnamefont{M.}~\bibnamefont{{Visser}}},
  \bibinfo{journal}{New Journal of Physics} \textbf{\bibinfo{volume}{6}},
  \bibinfo{pages}{186} (\bibinfo{year}{2004}), \eprint{arXiv:gr-qc/0408022}.

\bibitem[{\citenamefont{{Birrell} and {Davies}}(1984)}]{birreldavies}
\bibinfo{author}{\bibfnamefont{N.~D.} \bibnamefont{{Birrell}}}
  \bibnamefont{and} \bibinfo{author}{\bibfnamefont{P.~C.~W.}
  \bibnamefont{{Davies}}}, \emph{\bibinfo{title}{{Quantum Fields in Curved
  Space}}} (\bibinfo{publisher}{Cambridge University Press},
  \bibinfo{year}{1984}).

\bibitem[{\citenamefont{{Barcel{\'o}}
  et~al.}(2006{\natexlab{a}})\citenamefont{{Barcel{\'o}}, {Liberati}, {Sonego},
  and {Visser}}}]{particlecreation}
\bibinfo{author}{\bibfnamefont{C.}~\bibnamefont{{Barcel{\'o}}}},
  \bibinfo{author}{\bibfnamefont{S.}~\bibnamefont{{Liberati}}},
  \bibinfo{author}{\bibfnamefont{S.}~\bibnamefont{{Sonego}}}, \bibnamefont{and}
  \bibinfo{author}{\bibfnamefont{M.}~\bibnamefont{{Visser}}},
  \bibinfo{journal}{Classical and Quantum Gravity}
  \textbf{\bibinfo{volume}{23}}, \bibinfo{pages}{5341}
  (\bibinfo{year}{2006}{\natexlab{a}}), \eprint{arXiv:gr-qc/0604058}.

\bibitem[{\citenamefont{{Fulling} et~al.}(1978)\citenamefont{{Fulling},
  {Sweeny}, and {Wald}}}]{fsw}
\bibinfo{author}{\bibfnamefont{S.~A.} \bibnamefont{{Fulling}}},
  \bibinfo{author}{\bibfnamefont{M.}~\bibnamefont{{Sweeny}}}, \bibnamefont{and}
  \bibinfo{author}{\bibfnamefont{R.~M.} \bibnamefont{{Wald}}},
  \bibinfo{journal}{Communications in Mathematical Physics}
  \textbf{\bibinfo{volume}{63}}, \bibinfo{pages}{257} (\bibinfo{year}{1978}).

\bibitem[{\citenamefont{Simpson and Penrose}(1973)}]{Simpson:1973ua}
\bibinfo{author}{\bibfnamefont{M.}~\bibnamefont{Simpson}} \bibnamefont{and}
  \bibinfo{author}{\bibfnamefont{R.}~\bibnamefont{Penrose}},
  \bibinfo{journal}{Int. J. Theor. Phys.} \textbf{\bibinfo{volume}{7}},
  \bibinfo{pages}{183} (\bibinfo{year}{1973}).

\bibitem[{\citenamefont{{Poisson} and {Israel}}(1990)}]{poissonisrael}
\bibinfo{author}{\bibfnamefont{E.}~\bibnamefont{{Poisson}}} \bibnamefont{and}
  \bibinfo{author}{\bibfnamefont{W.}~\bibnamefont{{Israel}}},
  \bibinfo{journal}{\prd} \textbf{\bibinfo{volume}{41}}, \bibinfo{pages}{1796}
  (\bibinfo{year}{1990}).

\bibitem[{\citenamefont{{Markovi{\'c}} and {Poisson}}(1995)}]{markovicpoisson}
\bibinfo{author}{\bibfnamefont{D.}~\bibnamefont{{Markovi{\'c}}}}
  \bibnamefont{and}
  \bibinfo{author}{\bibfnamefont{E.}~\bibnamefont{{Poisson}}},
  \bibinfo{journal}{Physical Review Letters} \textbf{\bibinfo{volume}{74}},
  \bibinfo{pages}{1280} (\bibinfo{year}{1995}), \eprint{arXiv:gr-qc/9411002}.

\bibitem[{\citenamefont{{Macher} and {Parentani}}(2009)}]{parentani}
\bibinfo{author}{\bibfnamefont{J.}~\bibnamefont{{Macher}}} \bibnamefont{and}
  \bibinfo{author}{\bibfnamefont{R.}~\bibnamefont{{Parentani}}},
  \bibinfo{journal}{{to be published in \prd}}  (\bibinfo{year}{2009}),
  \eprint{{arXiv:hep-th/0903.2224}}.

\bibitem[{\citenamefont{{Barcel{\'o}} et~al.}(2005)\citenamefont{{Barcel{\'o}},
  {Liberati}, and {Visser}}}]{lrr-2005-12}
\bibinfo{author}{\bibfnamefont{C.}~\bibnamefont{{Barcel{\'o}}}},
  \bibinfo{author}{\bibfnamefont{S.}~\bibnamefont{{Liberati}}},
  \bibnamefont{and} \bibinfo{author}{\bibfnamefont{M.}~\bibnamefont{{Visser}}},
  \bibinfo{journal}{Living Reviews in Relativity} \textbf{\bibinfo{volume}{8}}
  (\bibinfo{year}{2005}), \eprint{(cited on \today)},
  \urlprefix\url{http://www.livingreviews.org/lrr-2005-12}.

\bibitem[{\citenamefont{{Barcel{\'o}}
  et~al.}(2006{\natexlab{b}})\citenamefont{{Barcel{\'o}}, {Liberati}, {Sonego},
  and {Visser}}}]{notrap}
\bibinfo{author}{\bibfnamefont{C.}~\bibnamefont{{Barcel{\'o}}}},
  \bibinfo{author}{\bibfnamefont{S.}~\bibnamefont{{Liberati}}},
  \bibinfo{author}{\bibfnamefont{S.}~\bibnamefont{{Sonego}}}, \bibnamefont{and}
  \bibinfo{author}{\bibfnamefont{M.}~\bibnamefont{{Visser}}},
  \bibinfo{journal}{Physical Review Letters} \textbf{\bibinfo{volume}{97}},
  \bibinfo{pages}{171301} (\bibinfo{year}{2006}{\natexlab{b}}),
  \eprint{arXiv:gr-qc/0607008}.

\end{thebibliography}

\end{document}